\documentclass[10pt,twocolumn,letterpaper]{article}

\usepackage{cvpr} 

\usepackage{graphicx}
\usepackage{amsmath}
\usepackage{amssymb}
\usepackage{booktabs}
\usepackage{float}

\usepackage{times}
\usepackage{epsfig}
\usepackage{amsmath}
\usepackage{multirow}
\usepackage{graphicx}

\usepackage{subcaption}
\newsavebox{\tempboxa}
\newsavebox{\tempboxb}
\newsavebox{\tempboxc}

\newcommand{\imgStubPDFpage}[4]{\begin{minipage}{#1\textwidth}\begin{center}
 \includegraphics[#2]{#3}\\
#4\end{center}\end{minipage}}

\newcommand{\Stress}[1]{\textbf{\textit{#1}}}
\newcommand{\wanyan}[1]{{#1}}
\newcommand{\xingbo}[1]{{#1}}

\usepackage[pagebackref,breaklinks,colorlinks]{hyperref}
\usepackage[accsupp]{axessibility} 

\usepackage[capitalize]{cleveref}
\crefname{section}{Sec.}{Secs.}
\Crefname{section}{Section}{Sections}
\Crefname{table}{Table}{Tables}
\crefname{table}{Tab.}{Tabs.}

\def\cvprPaperID{8572} 
\def\confName{CVPR}
\def\confYear{2022}

\begin{document}

\title{ Abandoning the Bayer-Filter to See in the Dark}

\author{Xingbo Dong\textsuperscript{1,3*}\footnotemark[2] ~ Wanyan Xu\textsuperscript{1,2*}\footnotemark[2] ~ Zhihui Miao\textsuperscript{1,2}\footnotemark[2] ~ Lan Ma\textsuperscript{1} ~ \\
Chao Zhang\textsuperscript{1} ~ Jiewen Yang\textsuperscript{1} ~ Zhe Jin\textsuperscript{4} ~ Andrew Beng Jin Teoh\textsuperscript{3} ~ Jiajun Shen\textsuperscript{1} \\
\textsuperscript{1}TCL Research AI Lab ~ \textsuperscript{2}Fuzhou University ~ \textsuperscript{3}Yonsei University ~ \textsuperscript{4}Anhui University\\
{\tt\small \{xingbo.dong,bjteoh\}@yonsei.ac.kr,\{208527051,208527090\}@fzu.edu.cn,\{sjj,rubyma\}@tcl.com}
}
\maketitle

\renewcommand{\thefootnote}{\fnsymbol{footnote}} 
\footnotetext[1]{These authors contributed equally to this work.} 
\footnotetext[2]{Work done while interning at TCL AI Lab.} 

\begin{abstract}
Low-light image enhancement - a pervasive but challenging problem, plays a central role in enhancing the visibility of an image captured in a poor illumination environment.
Due to the fact that not all photons can pass the Bayer-Filter on the sensor of the color camera, in this work, we first present a De-Bayer-Filter simulator based on deep neural networks to generate a monochrome raw image from the colored raw image. 
Next, a fully convolutional network is proposed to achieve the low-light image enhancement by fusing colored raw data with synthesized monochrome data. Channel-wise attention is also introduced to the fusion process to establish a complementary interaction between features from colored and monochrome raw images.
To train the convolutional networks, we propose a dataset with monochrome and color raw pairs named Mono-Colored Raw paired dataset (MCR) collected by using a monochrome camera without Bayer-Filter and a color camera with Bayer-Filter. 
The proposed pipeline take advantages of the fusion of the virtual monochrome and the color raw images and our extensive experiments indicate that significant improvement can be achieved by leveraging raw sensor data and data-driven learning. The project is available at \url{https://github.com/TCL-AILab/Erase_Bayer-Filter_to_See_in_the_Dark}
\end{abstract}
\renewcommand{\thefootnote}{\arabic{footnote}}
\section{Introduction}
For a digitalized image, the quality of the image could be severely degraded due to the color distortions and noise under poor illumination conditions such as indoors, nighttime, or improper camera exposure parameters.

\begin{figure}
\imgStubPDFpage{0.49}{page=2,width=0.99\linewidth,trim=0cm 2.8cm 9.8cm 0cm, clip}{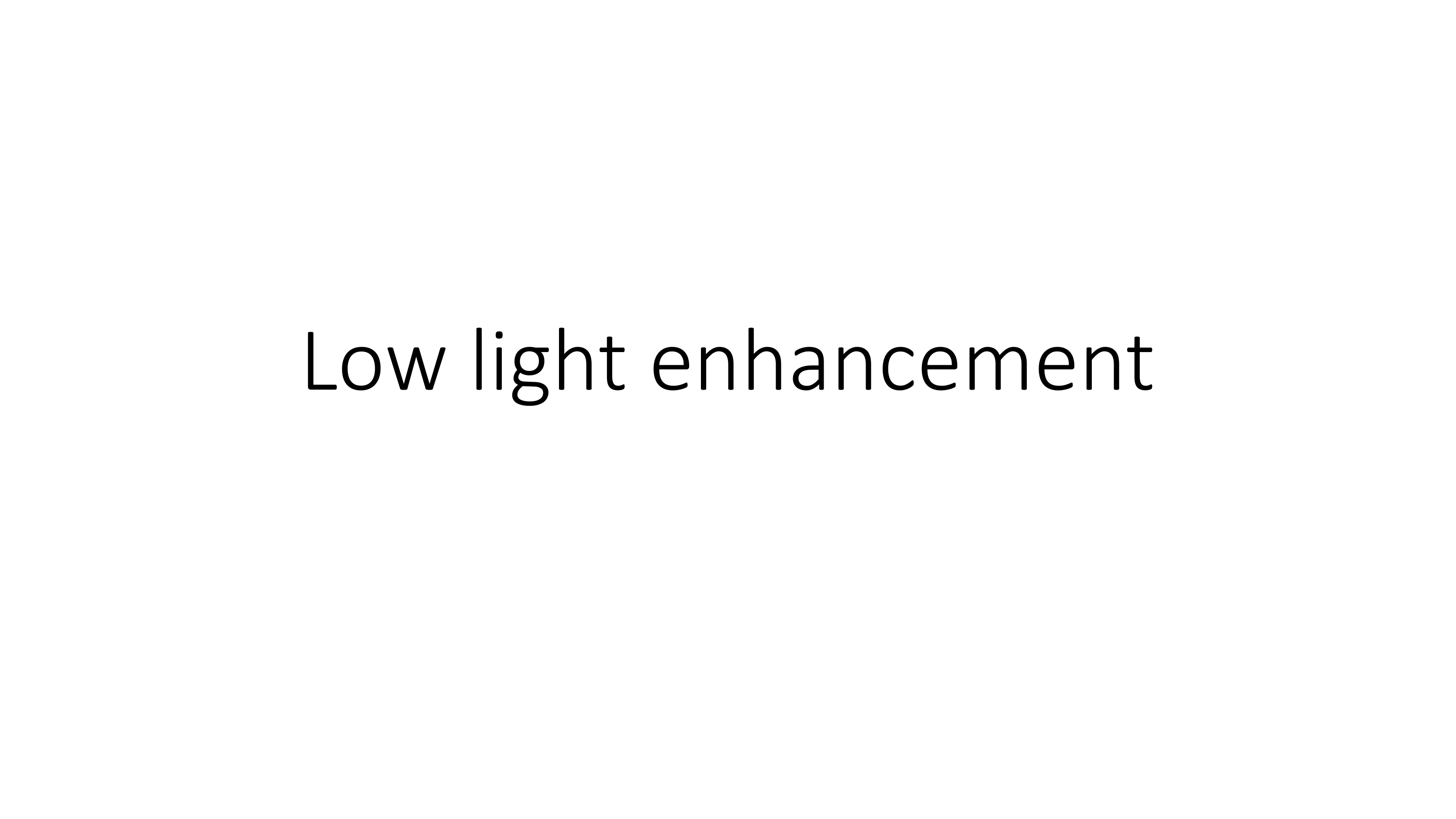}{} 
\caption{Overview of the proposed pipeline. We propose to generate monochrome raw data by a learned De-Bayer-Filter module. Then, a dual branch neural network is designed to bridge monochrome and colored raw to achieve the low-light image enhancement task.\label{Figure:overview}}
\end{figure}
Long exposure time and high ISO (sensitivity to light) are often leveraged in low-light environments to preserve visual quality. However, overwhelming exposure leads to motion blur and unbalanced overexposing, and high ISO amplifies the noise. Though the camera's flash provides exposure compensation for the insufficient light, it is not suitable for long-distance shooting, and it also introduces color distortions and artifacts. 
On the other hand, various algorithms have been reported to enhance the low-light image. Recently, deep neural network models have been utilized to solve the low-light image restoration \xingbo{problem}, such as DeepISP\cite{schwartz2018deepisp} and seeing in the dark (SID)\cite{chen2018learning}.

However, those algorithms are restricted in the image processing pipeline, as the photons capture rate and quantum efficiency are usually overlooked. In general, high photons capture rate can improve the image's visual quality significantly. One of the typical examples is the RYYB-based color filter, which can capture 40\% more photons than the Bayer-RGGB-based color filter\footnote{Bayer filter, Bayer-array, Bayer-array filter are used interchangeably. }. Hence, the RYYB-based color filter can achieve better performance naturally. 

\begin{figure*}[t]
\centering
\imgStubPDFpage{0.79}{page=1,width=0.99\linewidth,trim=0cm 11.8cm 1cm 0cm, clip}{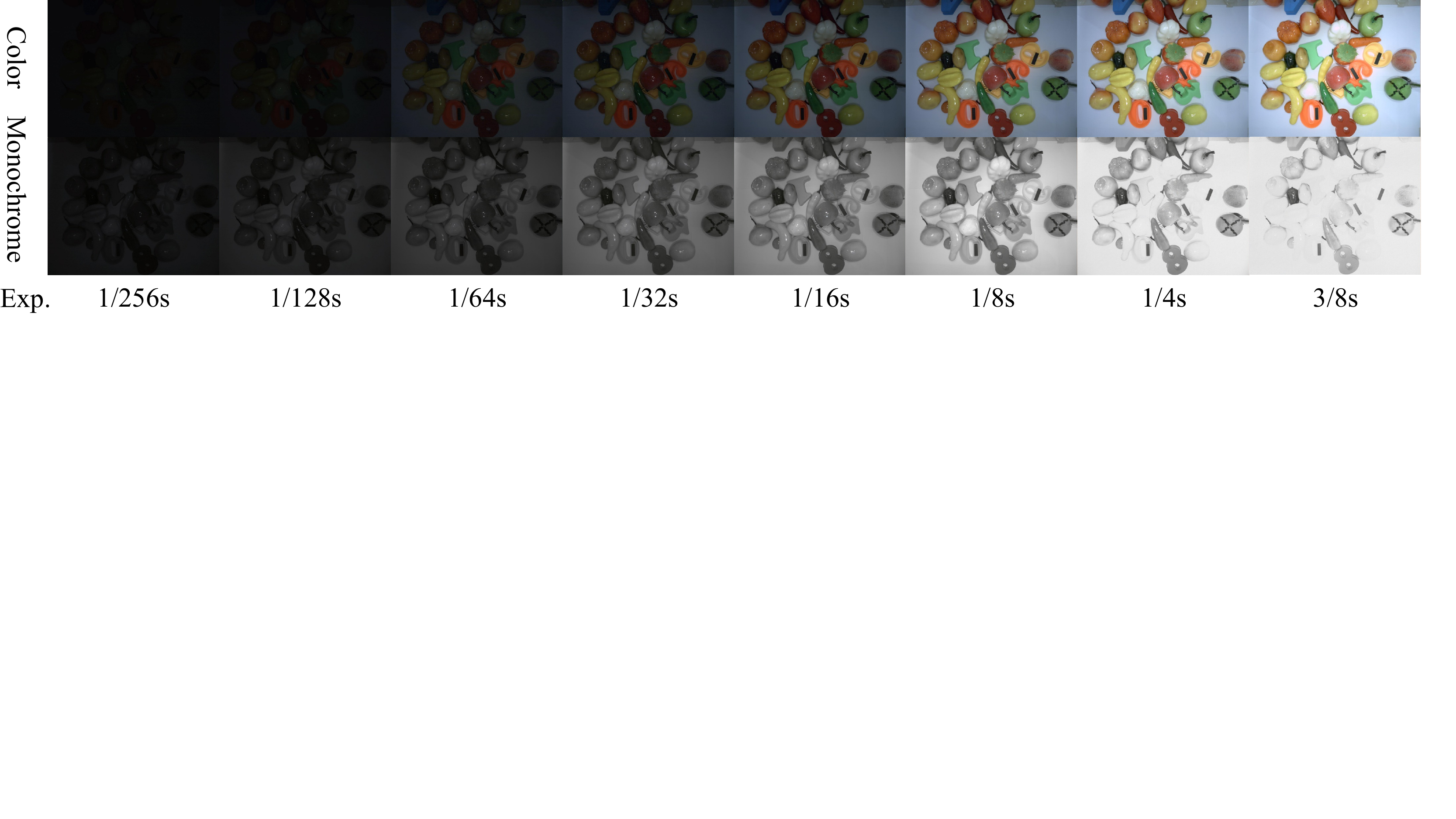}{(a)} 
\imgStubPDFpage{0.20}{page=7,width=0.70\linewidth,trim=0cm 11.8cm 28cm 0cm, clip}{diagram/diagram.pdf}{(b)} 
\caption{ (a) Images captured by color and monochrome cameras \xingbo{under different exposure time.}; (b) Monochrome and color cameras used in our work \xingbo{for data collection}. 
\label{Figure:Bayercamera}}
\end{figure*}

Bayer filter removal is another plausible way to improve the photons capture rate. The Bayer filter is an array of many tiny color filters that cover the image sensor to render color information (see Figure \ref{Figure:overview}). By removing the Bayer filter and sacrificing the color information, the image sensor can capture more photons, which contributes to \xingbo{clearer visibility} under poor illumination conditions \xingbo{compared with the camera with a Bayer filter} (see Figure \ref{Figure:Bayercamera} (a)). On the other hand, dual-cameras are one of the trends of today's smart devices such as smartphones. One type of dual-camera set is the combination of monochrome sensor and colored sensor\footnote{For example, Huawei P9, Moto Z2 Force}. The monochrome sensor is usually identical to the colored sensor but without a Bayer array filter. Such a dual-camera setting can achieve better imaging quality in a low-light environment due to more photons received by the sensor. However, additional cost is needed for the extra camera equipped. Therefore, for most mobile phones that are only equipped with color cameras, preserving the same low-light image quality produced by dual-camera set while only using a single color camera is a challenge task.

Motivated by the above discussion, we proposed a fully end-to-end convolutional neural model that consists of two modules (as illustrated in Figure \ref{Figure:overview}): a De-Bayer-Filter (DBF) module and a Dual Branch Low-light Enhancement module (DBLE). The DBF module learns to restore the monochrome raw image from the color camera raw data without requiring a monochrome camera. DBLE is designed to fuse colored raw with synthesized monochrome raw data and generate enhanced RGB images.

In addition, we propose a dataset to train our end-to-end framework. To the best of our knowledge, no existing dataset contains monochrome and colored raw image pairs captured by an identical type of sensors. To establish such a dataset, one camera with a Bayer filter is used to capture color-patterned raw images. Another camera without a Bayer-filter but equipped with the same type of sensor is utilized to capture monochrome raw images (see Figure \ref{Figure:Bayercamera} (b)). The dataset is collected \xingbo{under} various scenes, and each colored raw image has a corresponding monochrome raw image captured with identical exposure settings. 


Our contributions can be summarised as:
\begin{enumerate}
 \item A De-Bayer-Filter model is proposed to simulate a virtual monochrome camera and synthesize monochrome raw image data from the colored raw input.\wanyan{The DBF module aims at predicting the monochrome raw images, which resembles a monochrome sensor capability.}  To the best of our knowledge, we are the first to explore removing the Bayer-filter using a deep learning based model. 
 \item We design a Dual Branch Low-light Enhancement model that is used to fuse the colored raw with the synthesized monochrome raw to produce the final monitor-ready RGB images. To bridge the domain gap between colored raw and monochrome raw, a channel-wise attention layer is adopted to build an interaction between both domains for better restoration performance. The experiment results indicate that state-of-the-art performance can be achieved. 
 \item We propose the \textbf{MCR}, a dataset of colored raw and monochrome raw image pairs, captured with the same exposure setting. It is publicly opened as a research material to facilitate community utilization \xingbo{and will be released after publication}.
\end{enumerate}

\section{Related work}
To achieve the low-light image enhancement task, tremendous methods have been attempted. These methods can be categorized as histogram equalization (HE) methods \cite{lee2013contrast,wu2017contrast,arici2009histogram}, Retinex methods\cite{wei2018deep,zhang2019kindling,wang2019underexposed,fan2020integrating}, defogging model methods\cite{dong2011fast}, statistical methods \cite{su2017low,liang2015contrast,li2018restoration}, and machine learning methods\cite{xu2020learning,guo2020zero,kim2019low,zhu2020eemefn}. Recently, several works on raw image data have been proposed \cite{schwartz2018deepisp,chen2018learning,jiang2019learning}. Our work also falls into this category;  we will mainly discuss the existing methods of raw-based approaches in this section. 

Deep neural network become an emerging approach to achieve the digital camera's image signal processing tasks. 
In 2018, a fully convolutional model, namely DeepISP, was proposed in \cite{schwartz2018deepisp} to learn a mapping from the raw low-light mosaiced image to the final RGB image with high visual quality. To simulate the digital camera's image signal processing (ISP) pipeline, deepISP first extracts low-level features and performs local modifications, then extracts higher-level features and performs a global correction. L1 norm and the multi-scale structural similarity index (MS-SSIM) loss in the Lab domain are utilized for training the deepISP \xingbo{to simulate} the ISP pipeline. When DeepISP is only used for low-level imaging tasks such as denoising and demosaicing, L2 loss will be utilized. Hence, both low-level tasks and higher-level tasks such as demosaicing, denoising, and color correction can be achieved by DeepISP. The results in \cite{schwartz2018deepisp} suggest superior performance compared with manufacturer ISP.

Another parallel work similar to deepISP, namely seeing in the dark (SID), was proposed in \cite{chen2018learning}. In SID, a U-net \cite{ronneberger2015unet} network is utilized to operate directly on raw sensor data and outputs human visual ready RGB images. A dataset of raw short-exposure low-light images with corresponding long-exposure reference images was established to train the model. Compared with the traditional image processing pipeline, significant improvement can be made as the results in \cite{chen2018learning} indicate. Later, an improved version of SID was proposed in \cite{wang2019enhancement}. Using a similar u-net network as the backbone, the authors introduced wavelet transform to conduct down-sampling and up-sampling operations. Perceptual loss \cite{johnson2016perceptual} is used in \cite{wang2019enhancement} to train the network to better restore details in the image. In DID\cite{maharjan2019DID}, the authors proposed replacing the U-net in SID with residual learning to better preserve the information from image features. Similar raw-based approaches have also been applied to videos, such as \cite{jiang2019learning,chen2019seeing}.

\begin{figure*}[t]
\centering
\imgStubPDFpage{0.99}{page=6,width=0.93\linewidth,trim=5cm 21cm 7cm 5cm, clip}{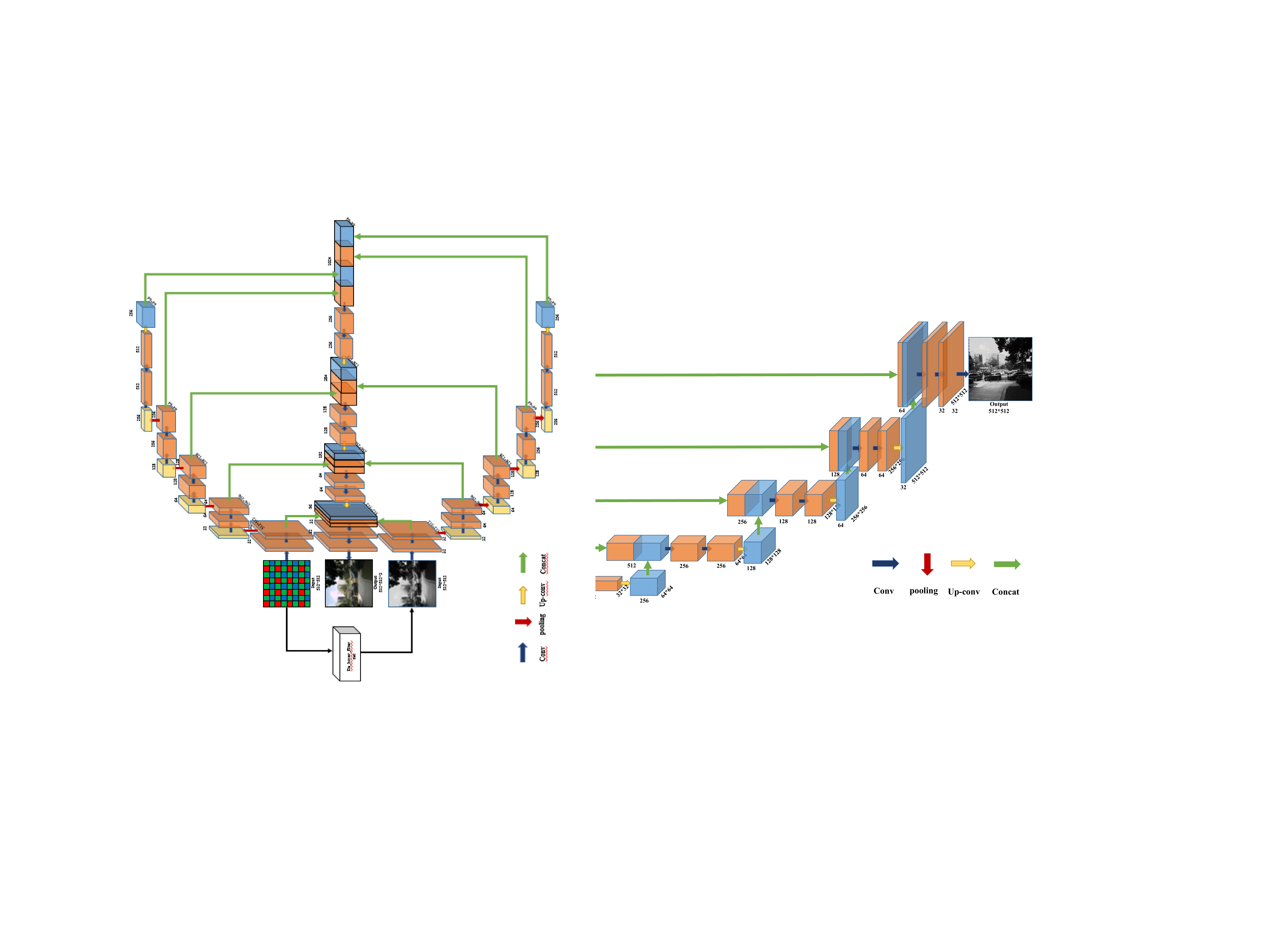}{(a) Architecture of the pipeline} 

\imgStubPDFpage{0.19}{width=0.99\linewidth}{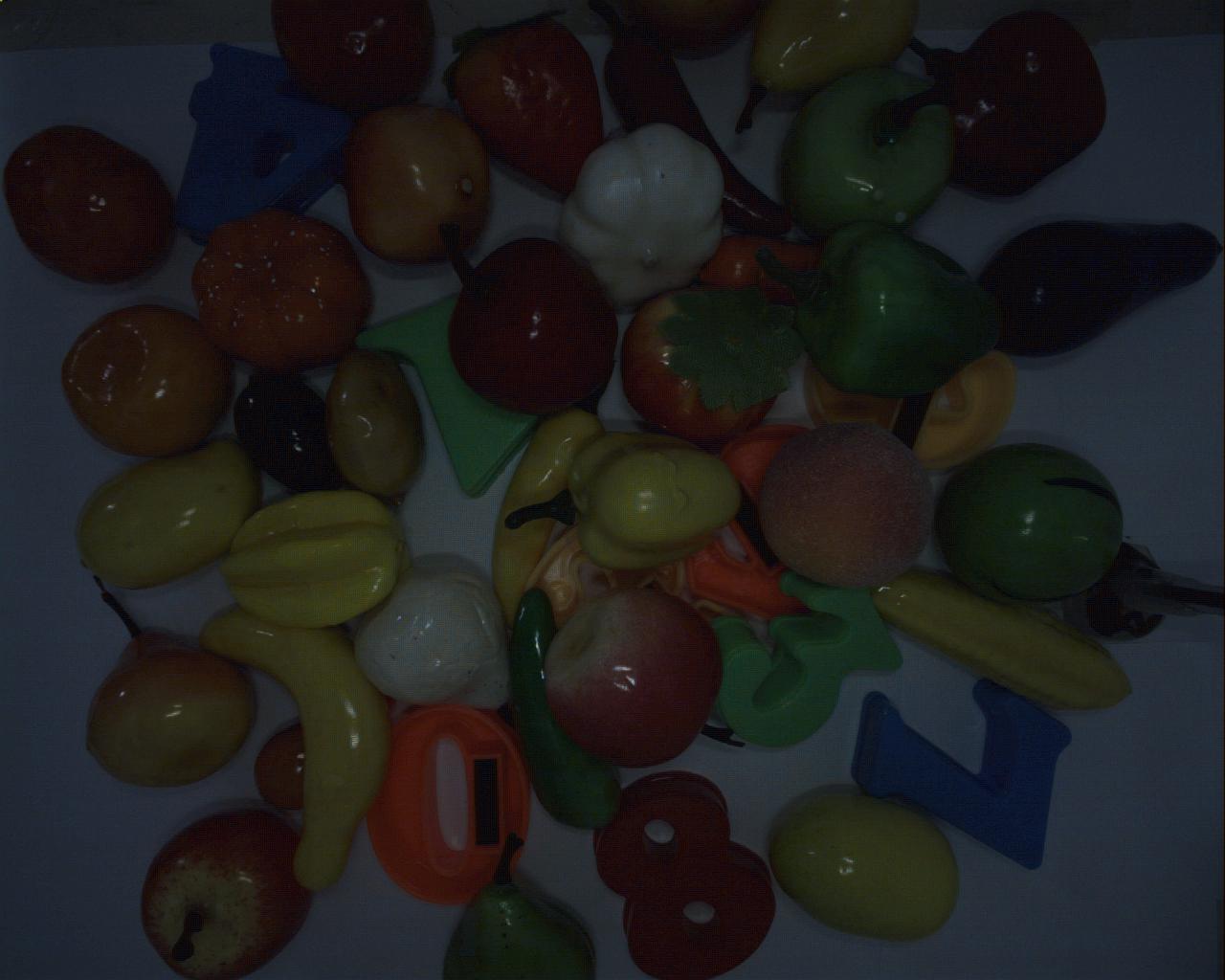}{(b) Input}
\imgStubPDFpage{0.19}{width=0.99\linewidth}{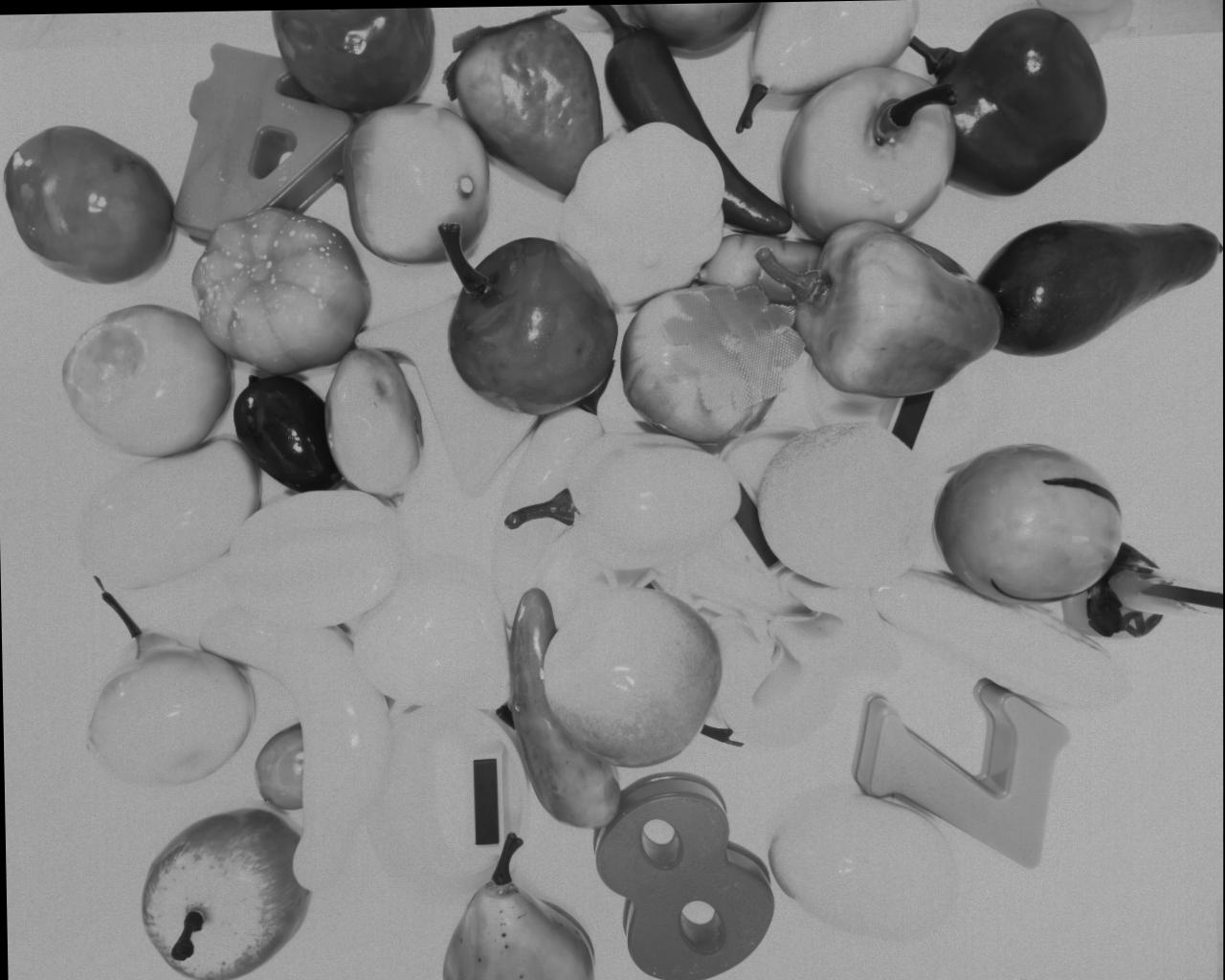}{(c) Mono GT}
\imgStubPDFpage{0.19}{width=0.99\linewidth}{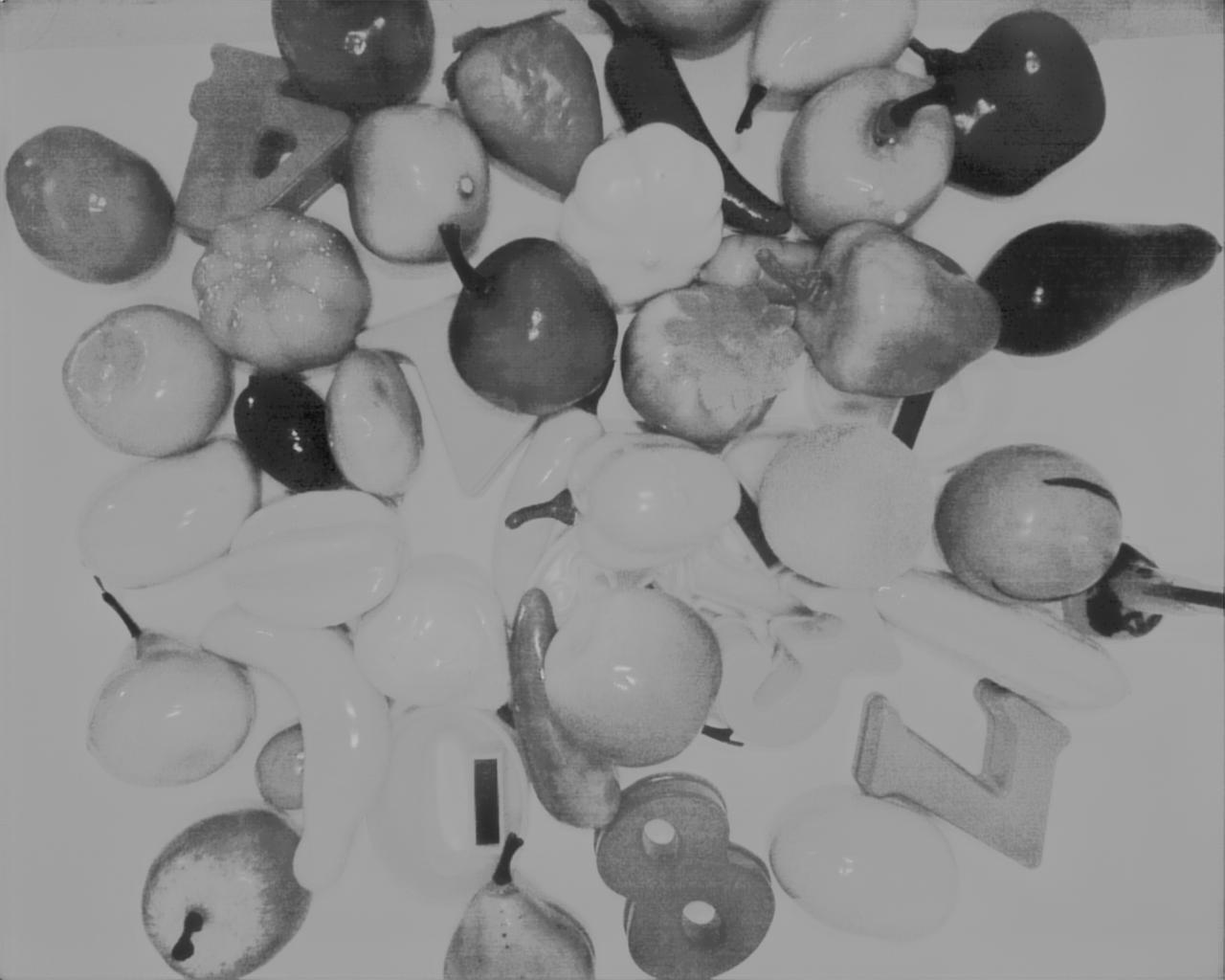}{(d) DBF output}
\imgStubPDFpage{0.19}{width=0.99\linewidth}{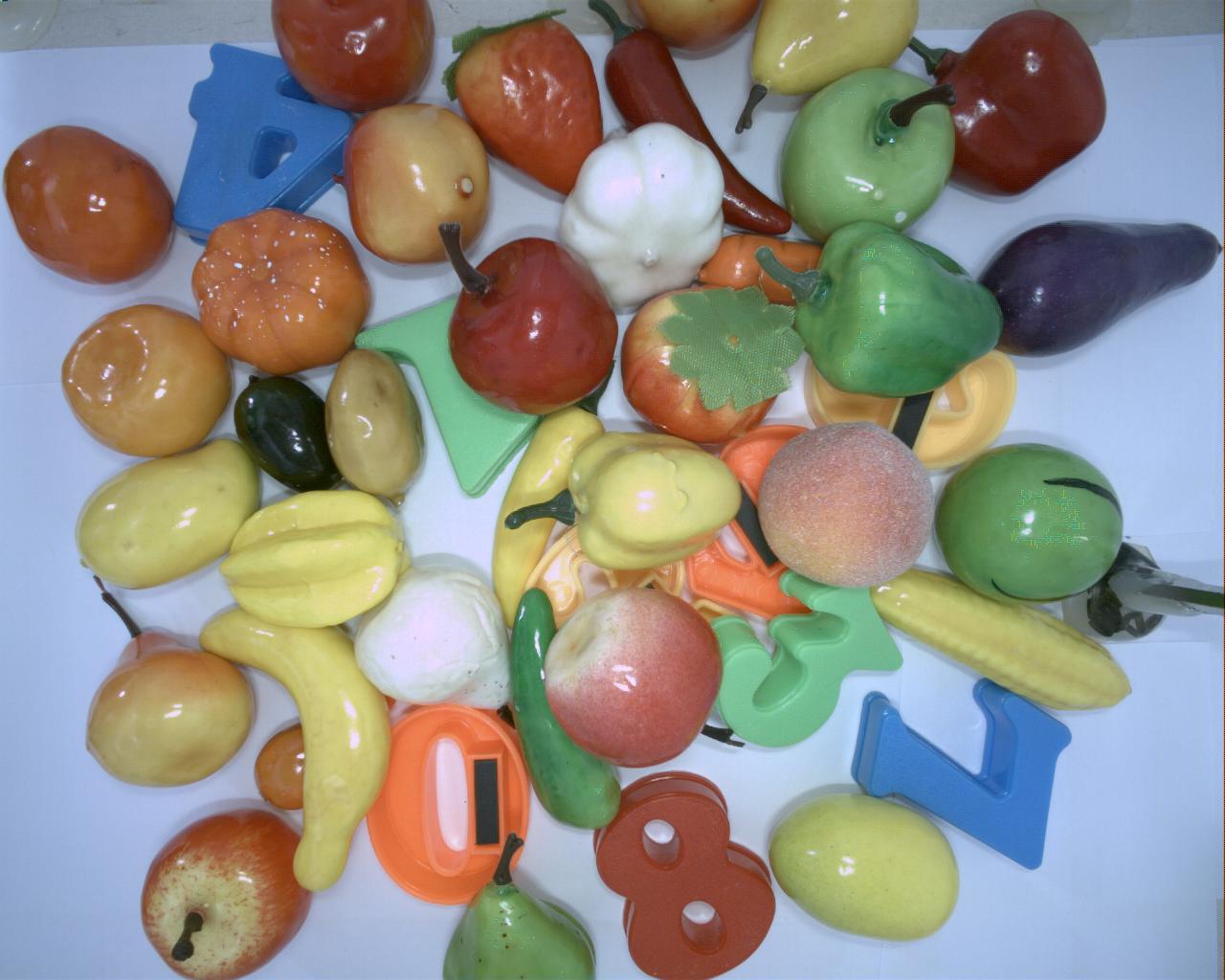}{(e) RGB GT}
\imgStubPDFpage{0.19}{width=0.99\linewidth}{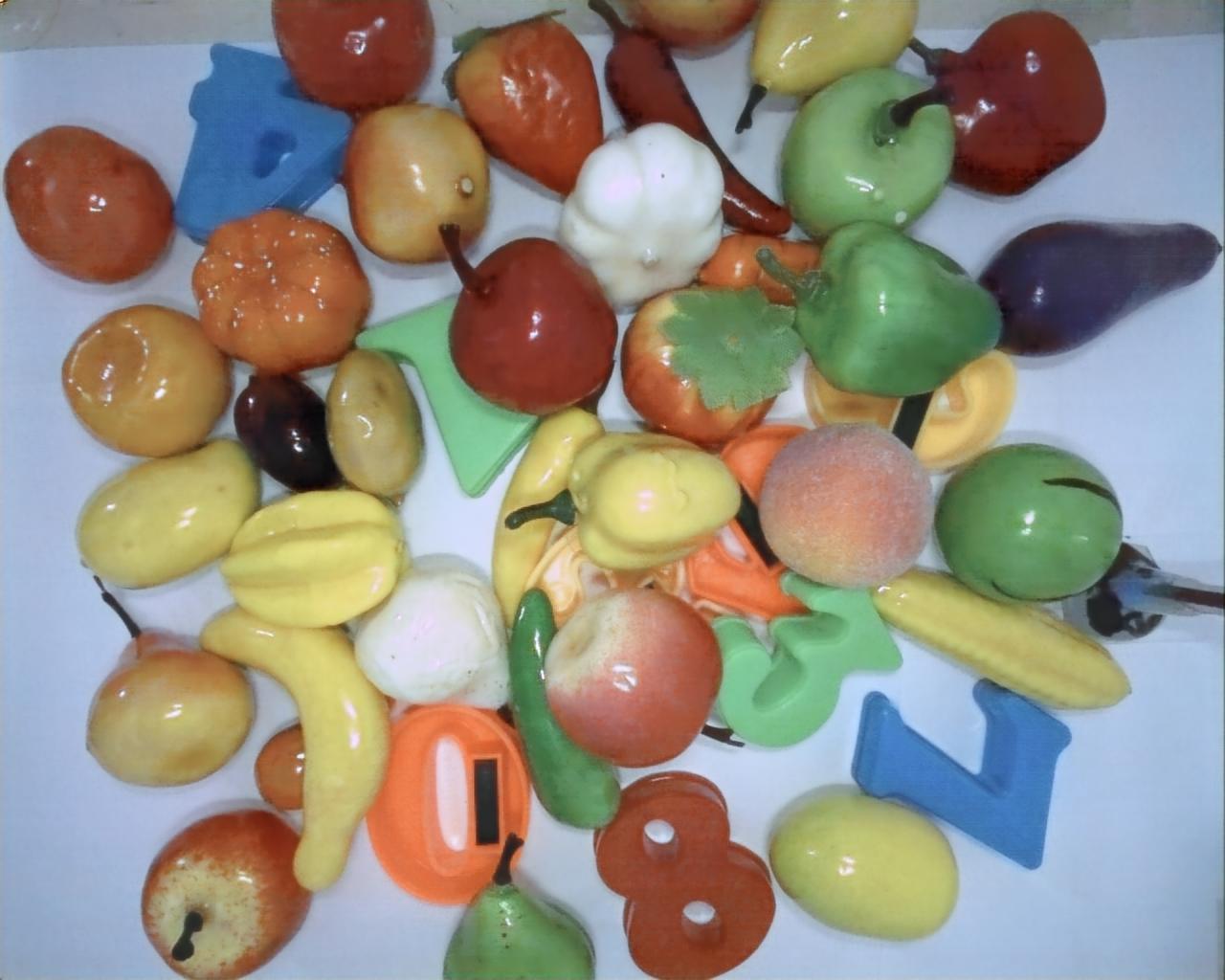}{(f) DBLE output}

\imgStubPDFpage{0.19}{width=0.99\linewidth}{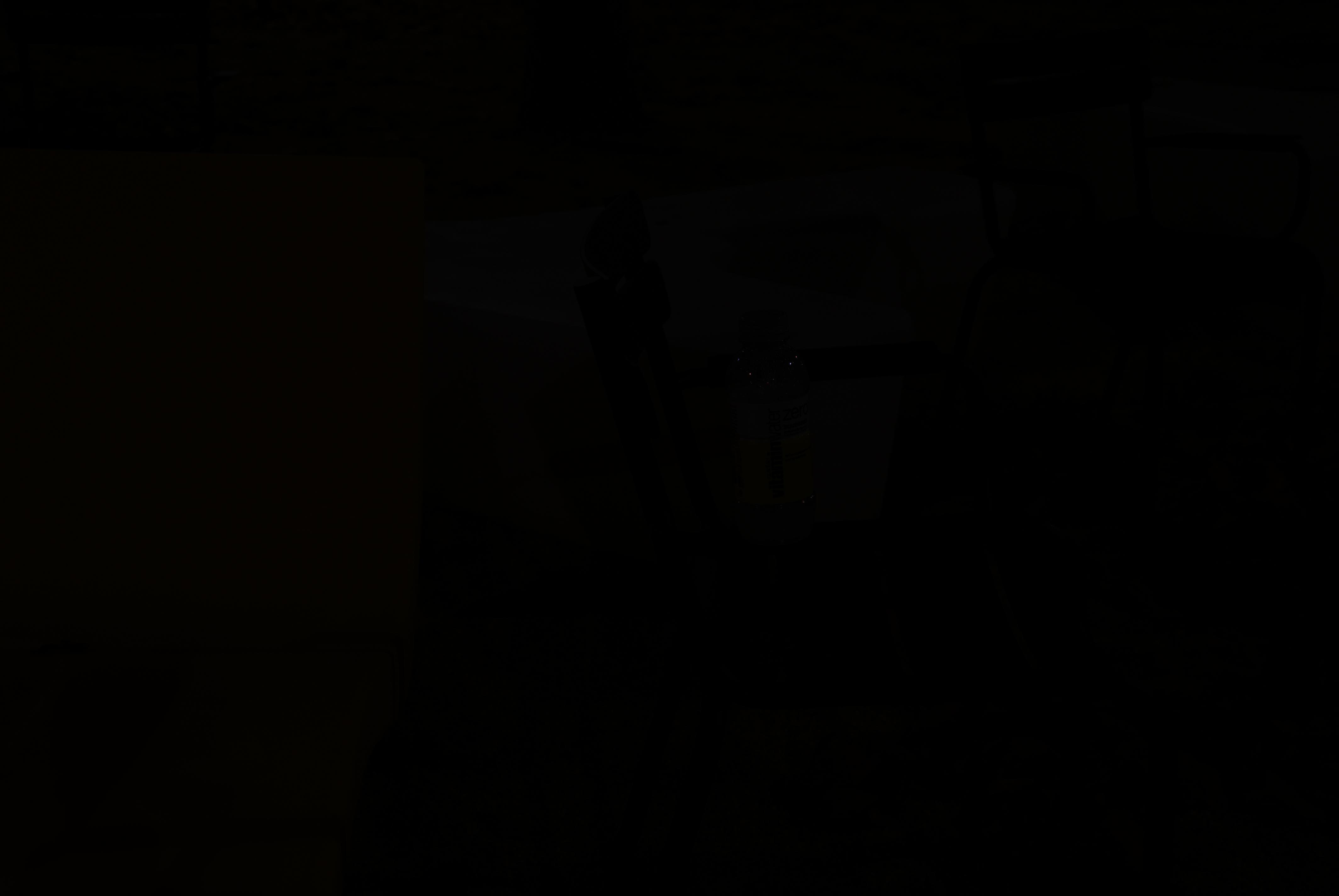}{(g) Input}
\imgStubPDFpage{0.19}{width=0.99\linewidth}{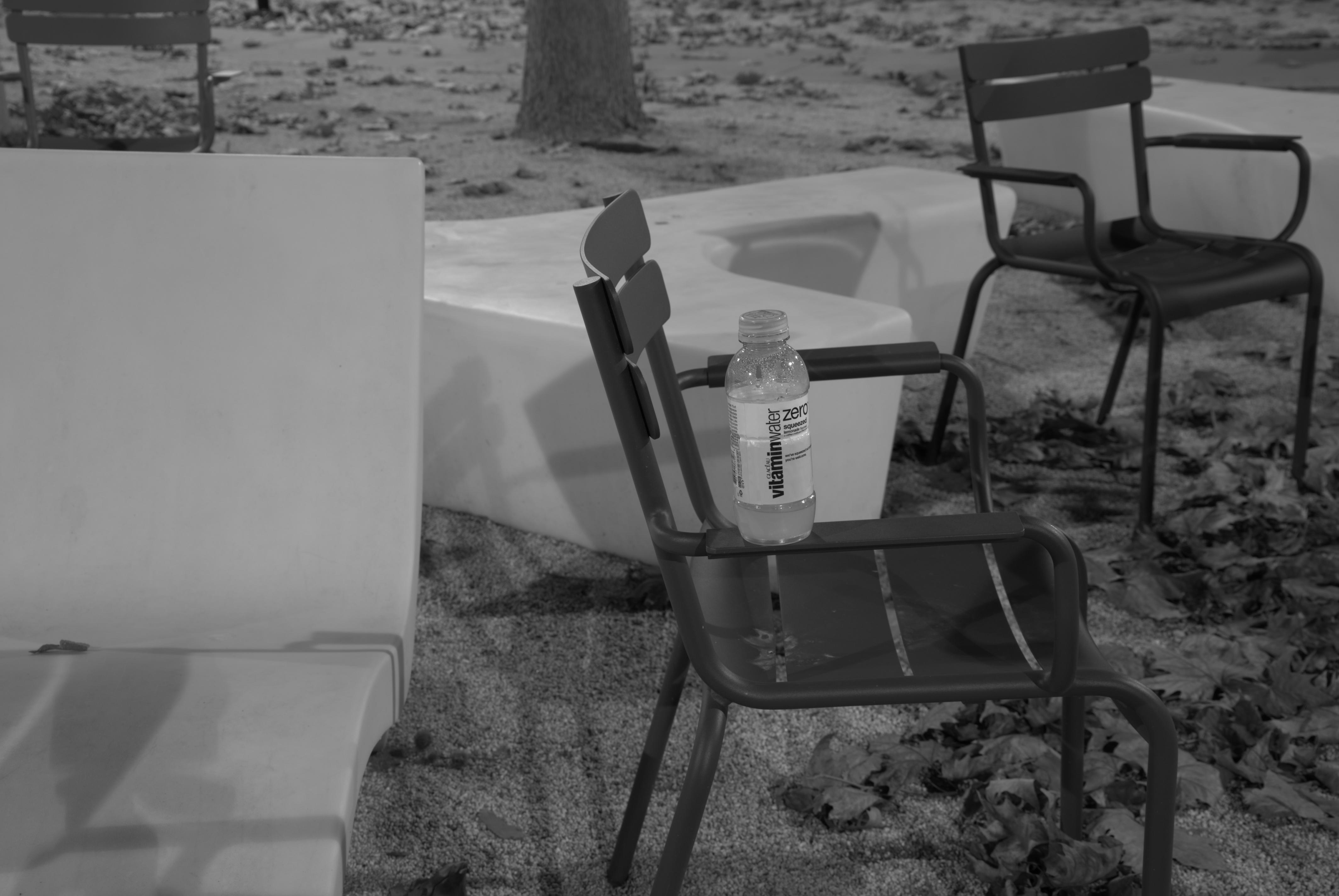}{(h) Synthetic Mono GT} 
\imgStubPDFpage{0.19}{width=0.99\linewidth}{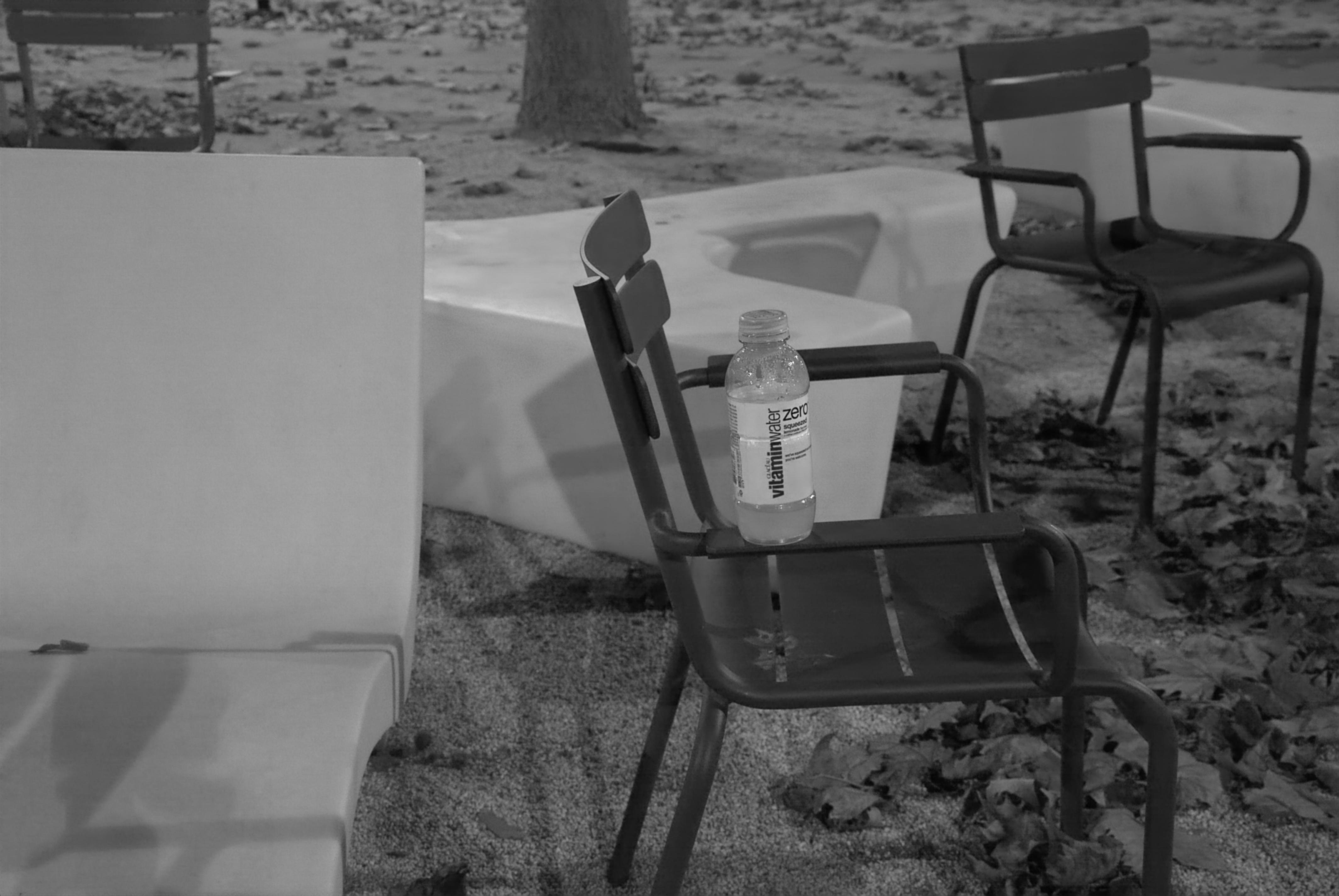}{(i) DBF output}
\imgStubPDFpage{0.19}{width=0.99\linewidth}{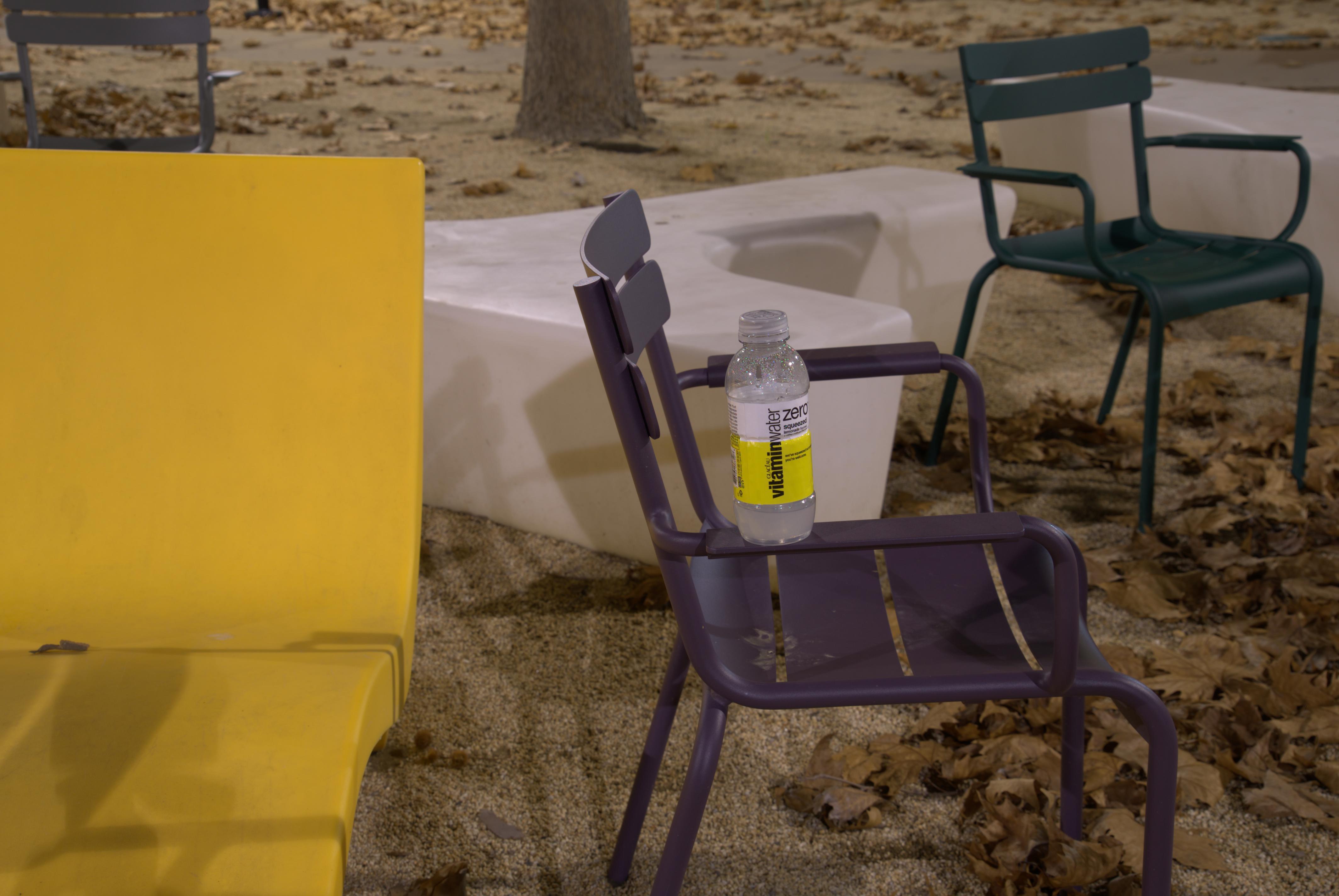}{(j) RGB GT}
\imgStubPDFpage{0.19}{width=0.99\linewidth}{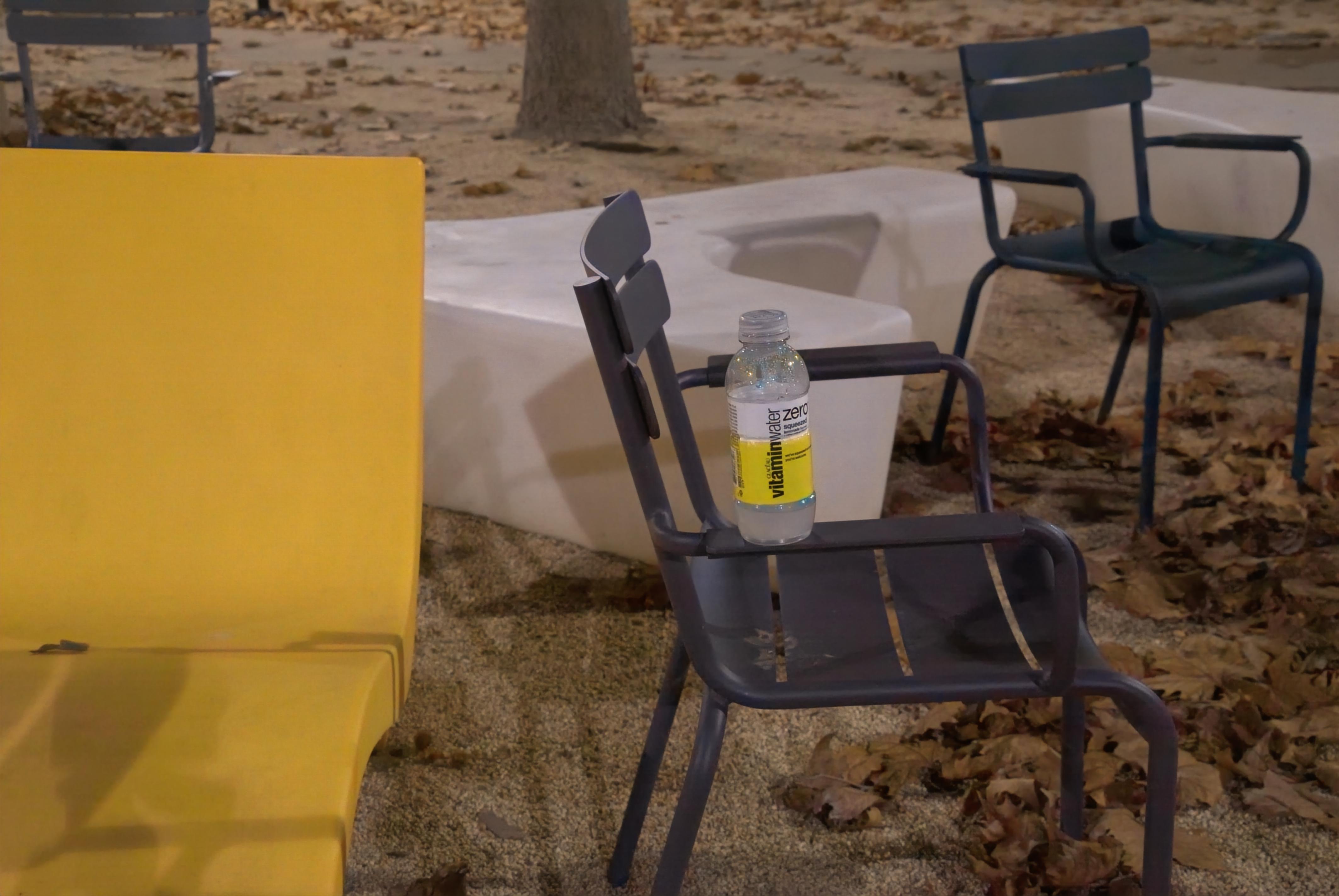}{(k) DBLE output}

\caption{(a) is the architecture of the pipeline. DBF module is designed to produce a monochrome image from the input raw image. DBLE module is proposed to fuse color and monochrome raw images to enhance the low-light input image. Each box denotes a multi-channel feature map produced by each layer. (b)-(f) are the images of our pipeline trained on our dataset. (g)-(k) are the images of our pipeline trained on SID\cite{chen2018learning} dataset, we convert RGB ground truth (GT) in SID dataset to gray image to replace the monochrome GT in our dataset.
\label{Figure:Virtualcameraoverview}}
\end{figure*}


On the top of the raw-based approach, frequency-based decomposition has also been explored on the low-light image enhancement task. 
In \cite{xu2020LDC}, the authors proposed a pipeline, namely LDC, to achieve the low-light image enhancement task based on frequency-based decomposition and enhancement model. The model first filters out high-frequency features and learns to restore the remaining low-frequency features based on an amplification operation. Subsequently, high-frequency details are restored. The results from \cite{xu2020LDC} indicate that state-of-the-art performance can be achieved by LDC.  


Several researches have also been done to improve the efficiency of low-light image enhancement in raw domain.
To achieve a computationally fast low-light enhancement system, the authors in \cite{RealTimeDarkImageRestorationCvpr2021} proposed a lightweight architecture (RED) for extreme low-light image restoration. Besides, the authors also proposed an amplifier module to estimate the amplification factor based on the input raw image. In \cite{gu2019SGN}, a self-guided neural network (SGN) was proposed to achieve a balance between denoising performance and the computational cost. It aims at guiding the image restoration process at finer scales by utilizing the large-scale contextual information from shuffled multi-resolution inputs.

Methods discussed above generally learn to map raw data captured by the camera to the human visual-ready image. As raw data provides full information, the reviewed approach achieves state-of-the-art performance. However, the performance of those methods is upper bounded by the information contained in the raw data. While in our work, we consider to introduce extra information beyond the raw-RGB data. 
\section{The method}
Motivated by the above discussion and inspired by the monochrome camera's high light sensitivity, we propose a novel pipeline to further push the raw-based approaches forward. Specifically, our pipeline takes a raw image captured by a color camera with Bayer-Filter as input. The De-Bayer-Filter module in our pipeline will first generate a monochrome image, and a dual branch low-light enhancement module then fuses the monochrome raw data and color raw data to produce the final enhanced RGB image. Both modules work on raw images, as raw images are linearly dependent on the number of photons received, which contains additional information  compared to RGB images such as the noise distribution \cite{chen2019seeing,ouyang2021neural}. Details of each module will be discussed \xingbo{in sequel}. A detailed architecture diagram of our framework is shown in Figure \ref{Figure:Virtualcameraoverview}(a) (more details are in supplementary). Furthermore, Figure \ref{Figure:Virtualcameraoverview}(b-f) and Figure \ref{Figure:Virtualcameraoverview}(g-k) visualize the output of each step of our model on our dataset and the SID dataset in \cite{chen2018learning}, respectively.

\subsection{De-Bayer-Filter module}
Millions of tiny light cavities are designed to collect photons and activate electrical signals on the camera sensor. However, using those light cavities alone can only produce gray images. A Bayer color filter is therefore designed to cover the light cavities and collect color information to produce color images. More specifically, a standard Bayer unit is a 2x2 pixel block with 2 green, 1 red and 1 blue color filter, and filters of a certain color will only allow photons with the corresponding wavelength to pass through. 

To simulate the camera imaging process using neural networks has been demonstrated feasible in several works \cite{schwartz2018deepisp,chen2018learning,ouyang2021neural}. Inspired by those works, we consider the removal of the Bayer array filter virtually by modeling the relationship between input and output photons for each color filter. Specifically, a De-Bayer-Filter (DBF) module is designed in this work to restore the monochrome raw images $A_{mono} \in \mathbb{R}^{H \times W}$ from the input colored raw $A_{ color} \in \mathbb{R}^{\frac{H}{2} \times \frac{W}{2} \times 4}$:
\begin{equation}
 A_{Mono} = f_{M}(A_{Color})
\end{equation}
where $f_{M}(\cdot)$ is a U-net-based fully convolutional network (see Figure \ref{Figure:Virtualcameraoverview}). \xingbo{L1 distance between the ground-truth monochrome image $A^{GT}_{Mono}$ and predicted image $A_{Mono}$ is used as a loss to encourage the DBF to learn to restore monochrome images with more details from low-light raw images.} We hypothesize that the generated monochrome raw image can 
enhance the low-light image by introducing more information into the subsequent module.
\subsection{Dual branch low-light image enhancement module}
There \xingbo{are} many differences between the colored raw image and monochrome image: 1) colored raw images have mosaic patterns; 2) the colored raw images consist of four channels with a resolution of $\frac{H}{2} \times \frac{W}{2}$, while its counterpart consists of one channel with $H \times W$ resolution; 3) no color information is included in the monochrome images; 4) better illuminating information is preserved on monochrome images as the monochrome camera sensor can better capture the light. 

Based on the above observations, we propose a dual branch low-light image enhancement (DBLE) module (see Figure \ref{Figure:Virtualcameraoverview}), which treats the DBF generated monochrome raw image and colored raw image separately in the down-sampling process. Meanwhile, different level feature maps of the two down-sampling branches are fused based on concatenation and followed by channel-wise attention (CA) layer\cite{hu2018squeeze} in the up-sampling branch to synthesize the human visual ready RGB image $I_{rgb} \in \mathbb{R}^{H \times W \times 3}$. The DBLE module is defined as:
\begin{equation}
 I_{RGB} = f_{C}(A_{Color};A_{Mono}),
\end{equation}
where $f_{C}$ is a specifically designed fully convolutional network, which is shown in Figure \ref{Figure:Virtualcameraoverview} (a). \xingbo{L1 distance between the ground truth RGB image $I_{RGB}^{GT}$ and predicted image $I_{RGB}$ is used as the loss to encourage the DBLE to learn to restore visual-ready RGB output from low-light raw images.} 


As the conventional U-net network treats features from each channel equally, directly concatenating the feature map from the monochrome raw branch and colored raw branch may lead to contradiction due to the domain gap. The usage of strided convolution and transposed convolution layer will also lead to spatial information loss. \xingbo{Motivated} by \cite{zhang2019cross}, after the concatenation operation, a CA layer \cite{hu2018squeeze} is adopted to achieve a channel-wise attention recalibration in DBLE to bridge the gap between monochrome and color images. The CA layer can explicitly model the interaction of colored raw and monochrome raw modalities to exploit the complementariness and reduce contradiction from both domains.

It has been reported that upsampling layers (transposed convolutional layers) used in U-net causes images to be distorted by checkerboard artifacts \cite{odena2016deconvolution,sugawara2018super,sugawara2019checkerboard,kinoshita2020fixed}. We also found such checkerboard artifacts in our settings on U-Net, especially for images with white background. In our work, the CA layer also serves a role to avoid checkerboard artifacts. As downscale and upscale operations are included in the CA layer, the CA layer is similar to the resize-convolution operation which discourages high-frequency artifacts in a weight-tying manner \cite{odena2016deconvolution}.


\subsection{Dataset design}
\Stress{Mono-colored raw Paired dataset (\textbf{MCR})}. To the best of our knowledge, no existing dataset contains monochrome and Bayer raw image pairs captured by the same type of sensors. 
To establish the dataset, we capture image pairs of the same scenes with two cameras, denoted as Cam-Color and Cam-Mono\footnote{Part Number: MT9M001C12STC/MT9M001C12STM}. Both cameras have the same 1/2-inch CMOS sensor and output a 1,280H x 1,024V imaging pixel array. However, only Cam-Color is equipped with a Bayer color filter. Cam-Color is used to capture colored raw images in our work, and Cam-Mono captures monochrome raw images.

We collect the data in both indoor and outdoor conditions. The illuminance at the indoor scenes is between 50 lux and 2000 lux under regular lights. The outdoor images were captured during daytime and night, under sun lighting or street lighting, with an illuminance between 900 lux and 14000 lux. The captured scene includes toys, books, stationery objects, street view, and parks. 

The cameras are mounted on the sliding platform on sturdy tripods or a fixed platform on a sturdy table. When mounted on the sliding platform, the camera is adjusted to the same position by sliding the platform to minimize the position displacement among images captured by two cameras in the same scene. When mounted on the fixed platform, the camera is attached to the same position of the platform to minimize the position displacement. Camera gain is set with the camera default value. Focal lengths are adjusted to maximize the quality of the images under long exposure. The exposure time is adjusted according to the specific scene environment.

Position displacement is unavoidable in the capture process. Hence, it is necessary to align the images captured from two cameras. The best exposure colored raw and monochrome raw is selected to align the images captured by two cameras in the same scenes. Then, homography feature matching is utilized to extract key points from the selected image pair, and a brute force matcher is utilized to find the matched key points. The extracted locations of good matches are filtered based on an empirical thresholding method. A homography matrix can be decided based on the filtered location of good matches. Finally, the homography transformation is applied to other images captured from the same scene. The statistic information of the dataset is summarized in Table \ref{table:ds_summary}. Figure \ref{Figure:Bayercamera}(a) demonstrates a series of monochrome-colored raw paired images from the dataset.

\begin{table}
\centering
\caption{Summary of the dataset \label{table:ds_summary}}
\resizebox{\linewidth}{!}{%
\begin{tabular}{l|c|l|l} 
\hline
Scenes & Exposure time (s) & Data Pairs & Fixed Settings \\ 
\hline
Indoor fixed position~ & \begin{tabular}[c]{@{}l@{}}1/256, 1/128, 1/64, 1/32,\\ 1/16, 1/8, 1/4, 3/8\end{tabular} & 2744  pairs & \multirow{3}{*}{\begin{tabular}[c]{@{}l@{}}Format: .raw,\\resolution: \\1280*1024~ ~ ~ ~ ~\end{tabular}} \\ 
\cline{1-3}
Indoor sliding platform & \begin{tabular}[c]{@{}l@{}}1/256, 1/128, 1/64, 1/32,\\ 1/16, 1/8, 1/4, 3/8\end{tabular} & 800 pairs & \\ 
\cline{1-3}
Outdoor sliding platform & \begin{tabular}[c]{@{}c@{}}1/4096, 1/2048, 1/1024, 1/512,\\1/256, 1/128, 1/64, 1/32\end{tabular} & 440 pairs & \\
\hline
\end{tabular}
}
\end{table}

\Stress{Artificial Mono-colored raw SID dataset}. 
The original SID dataset collected in \cite{chen2018learning} contains 5094 raw short-exposure images taken from the indoor and outdoor environment, while each short-exposure image has a corresponding long-exposure reference image. The short exposure time is usually between 1/30 second and 1/10 second, and the exposure time of the corresponding long-exposure image is 10 to 30 seconds. 

However, monochrome images are not available in the original SID dataset. To address this, we built an artificial Mono-colored raw dataset based on SID\cite{chen2018learning} dataset in this work. More specifically, we first convert the long-exposure raw images in the original SID dataset to RGB images, and these RGB images are further converted to grayscale by forming a weighted sum of the R, G, and B channels, as is shown in Figure \ref{Figure:Virtualcameraoverview}(h). Such conversion can eliminate the hue and saturation information while retaining the luminance information. 

\begin{figure*}
\centering
\imgStubPDFpage{0.19}{width=0.99\linewidth}{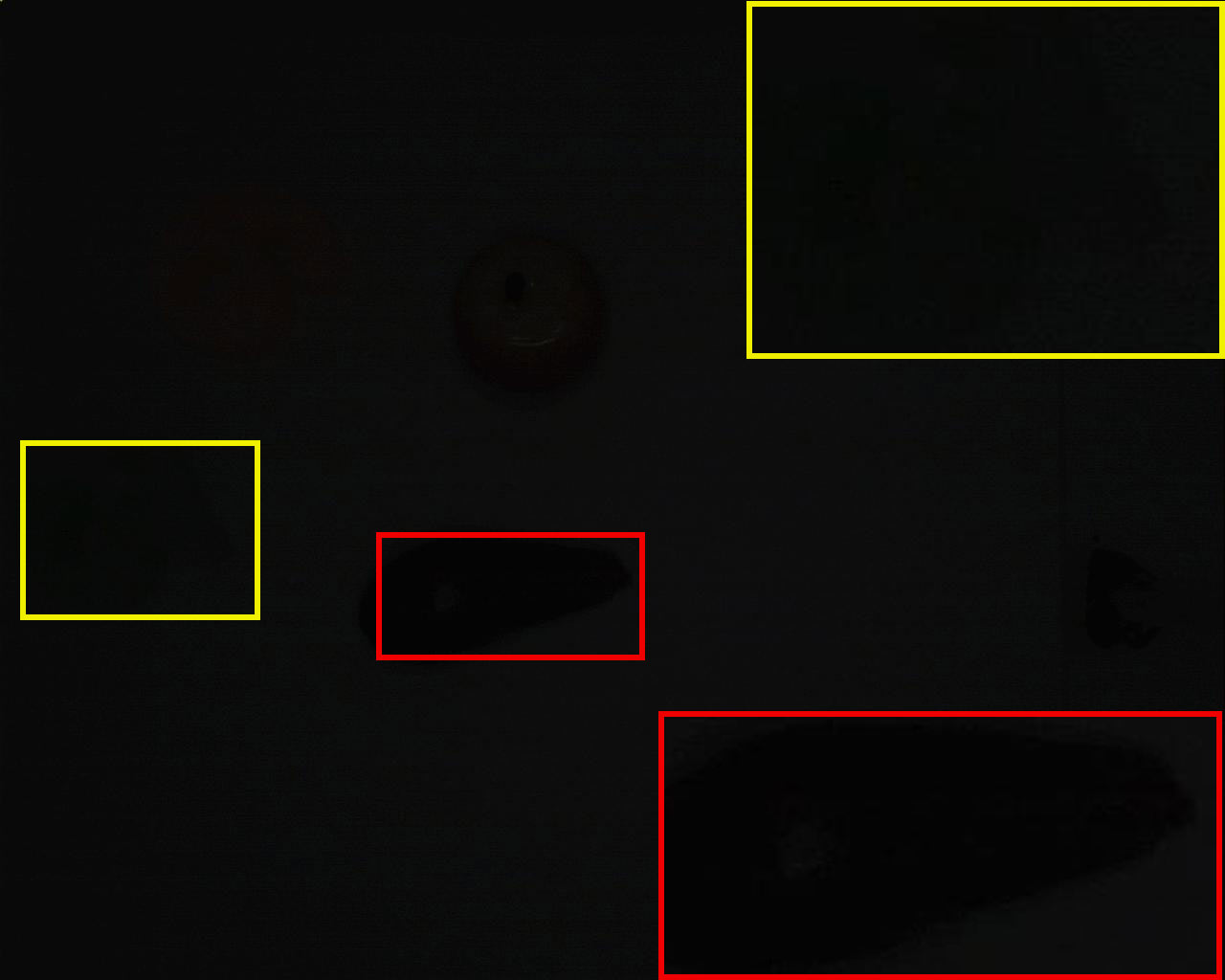}{(a) Input raw }
\imgStubPDFpage{0.19}{width=0.99\linewidth}{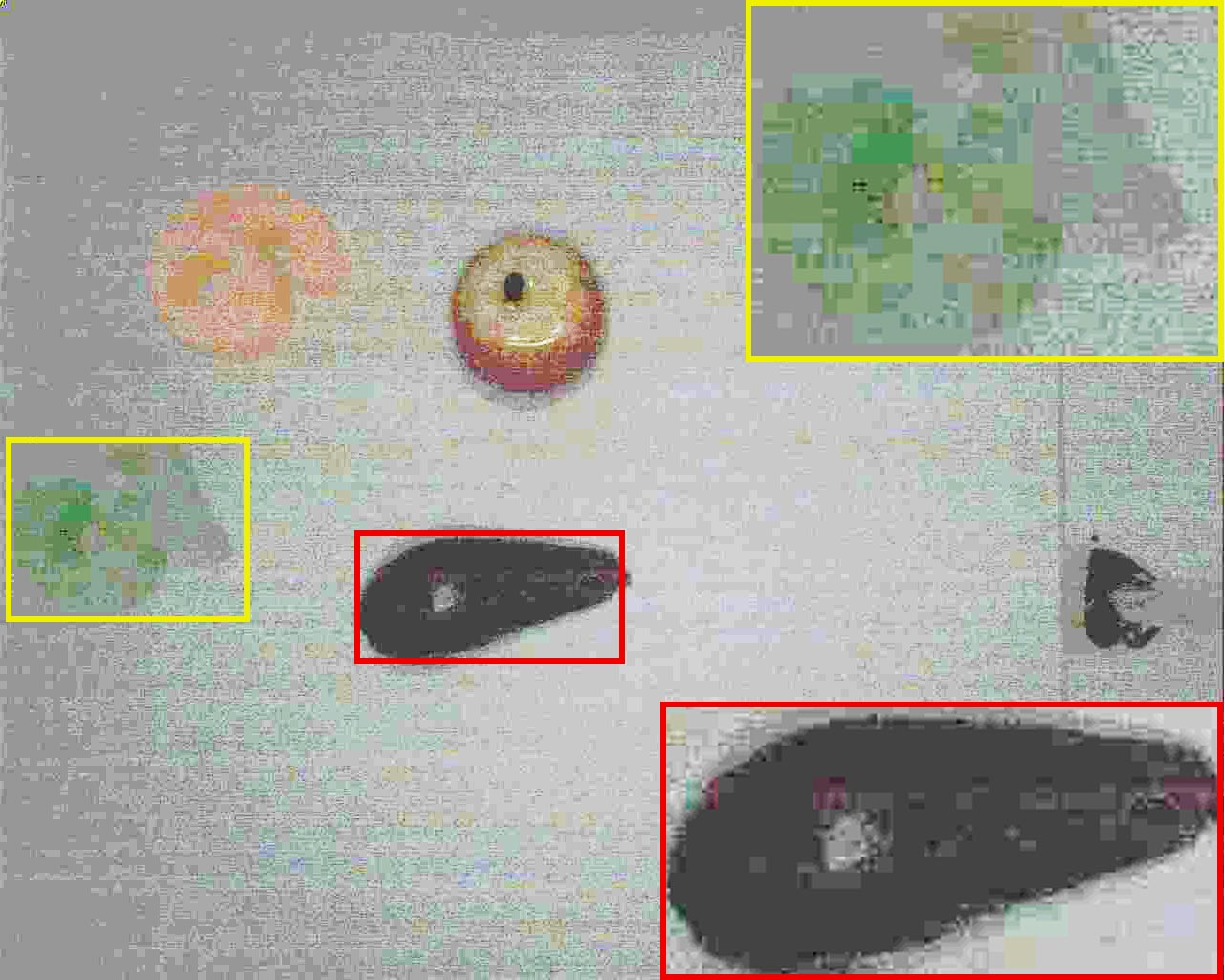}{(b) CSAIE}
\imgStubPDFpage{0.19}{width=0.99\linewidth}{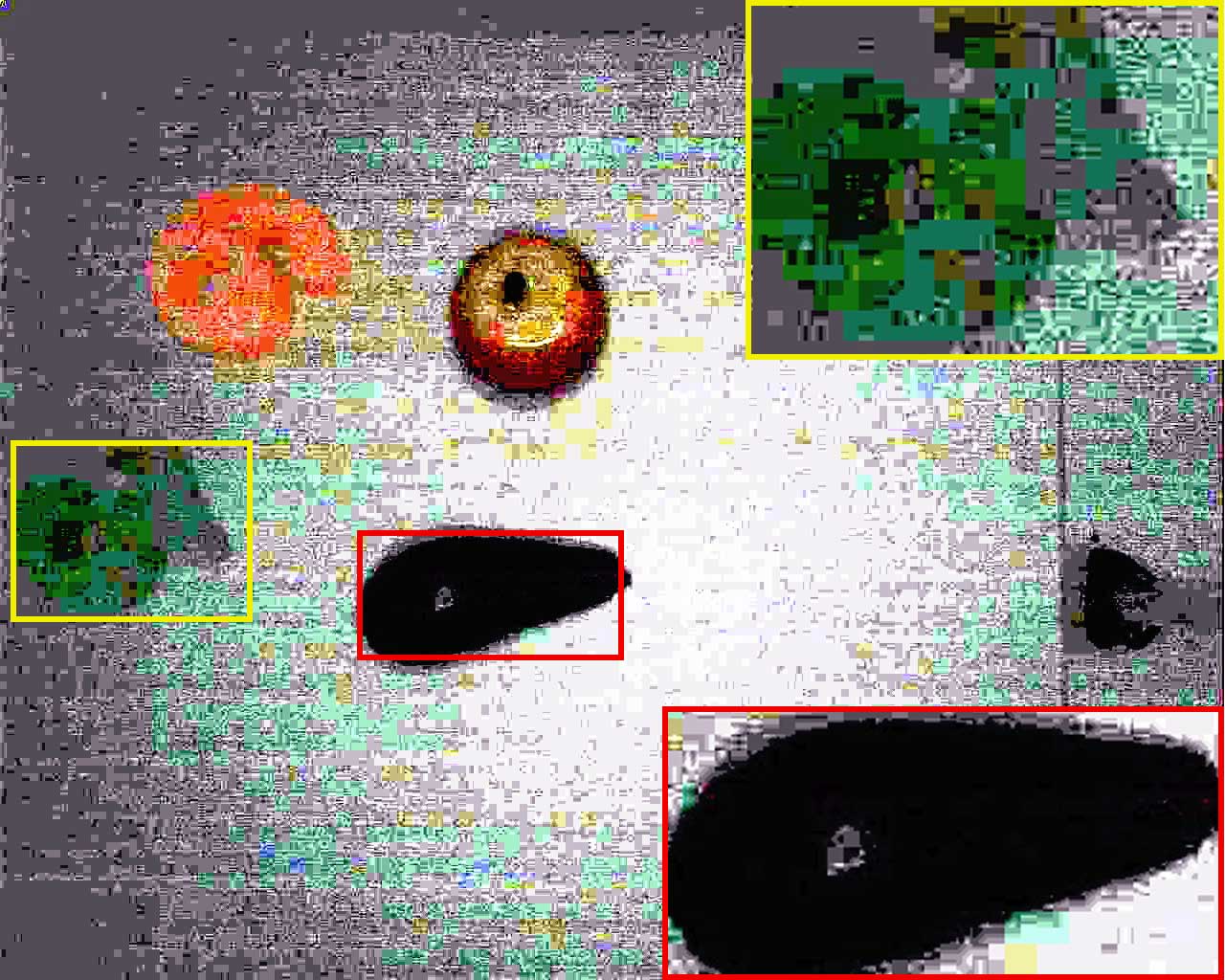}{(c) HE }
\imgStubPDFpage{0.19}{width=0.99\linewidth}{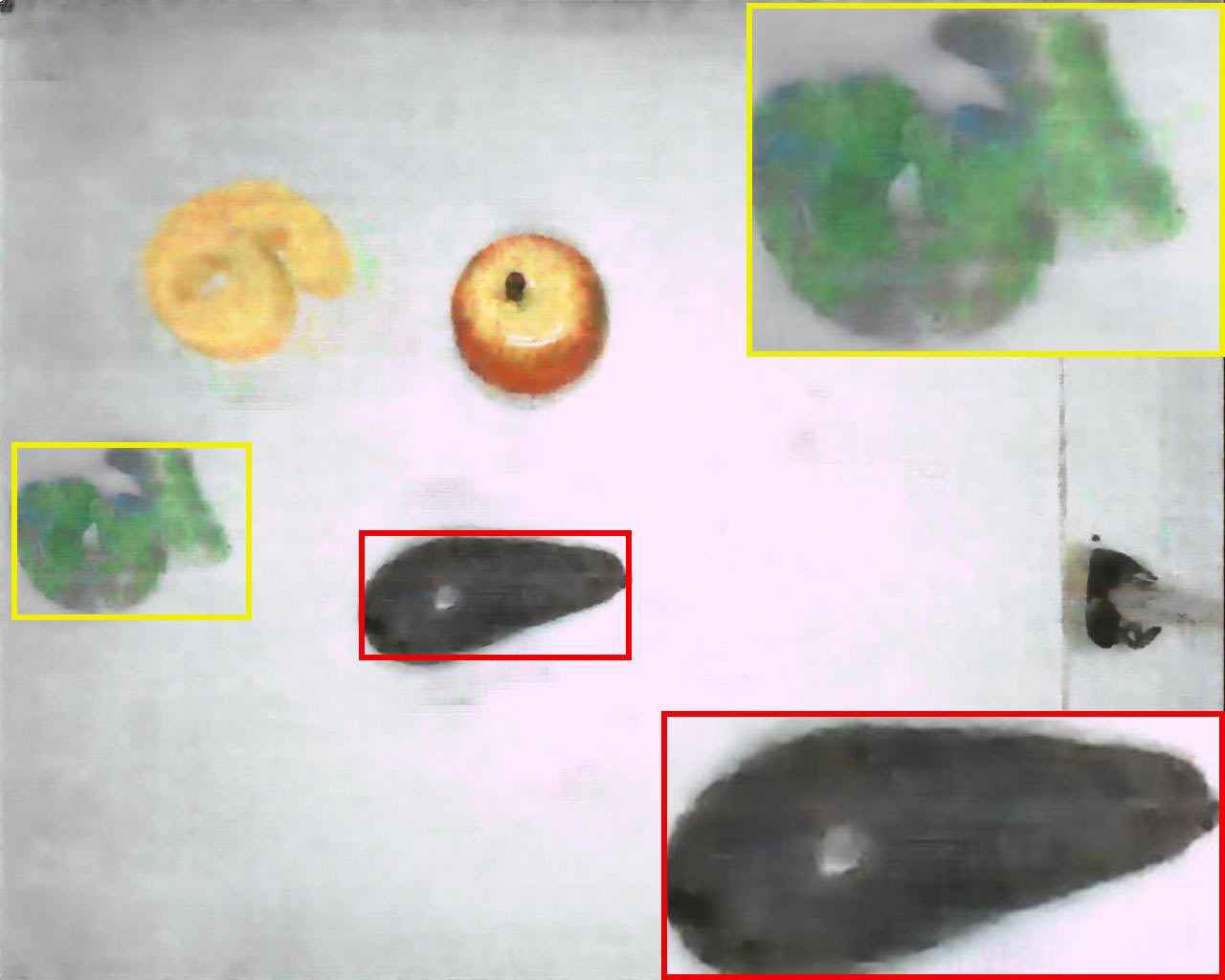}{(d) SGN \cite{gu2019SGN} }
\imgStubPDFpage{0.19}{width=0.99\linewidth}{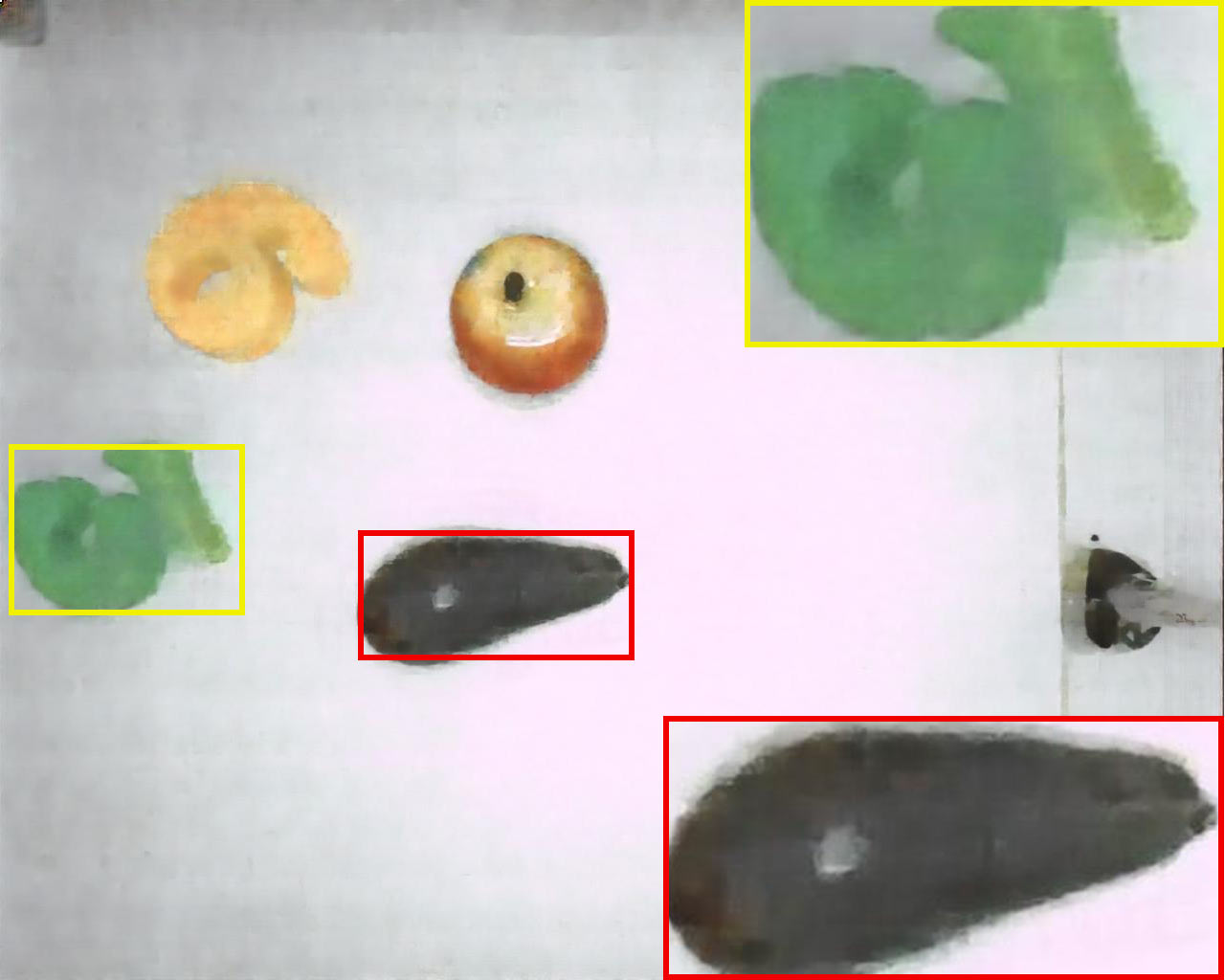}{(e) DID\cite{maharjan2019DID} }
\imgStubPDFpage{0.19}{width=0.99\linewidth}{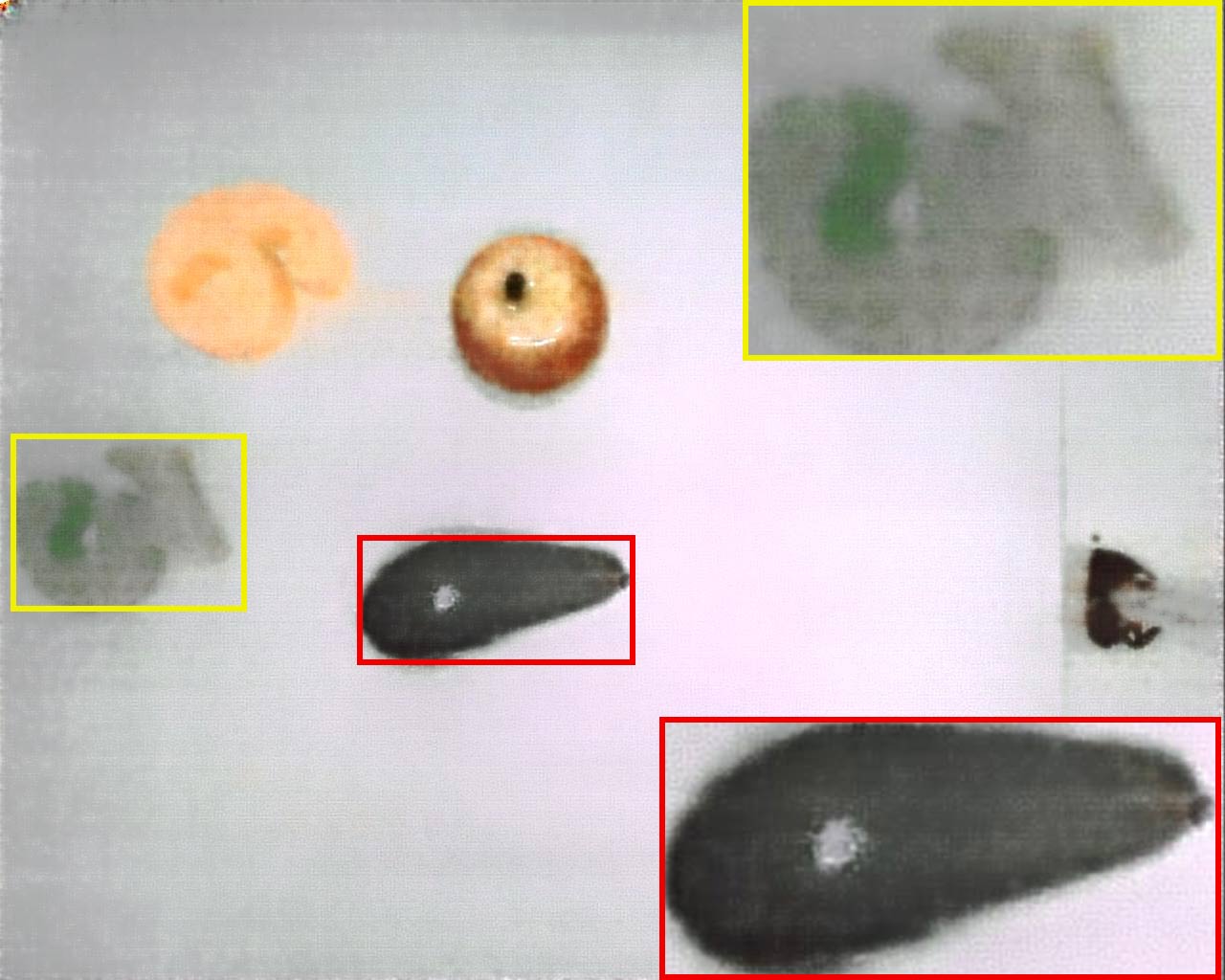}{(f) RED\cite{RealTimeDarkImageRestorationCvpr2021} }
\imgStubPDFpage{0.19}{width=0.99\linewidth}{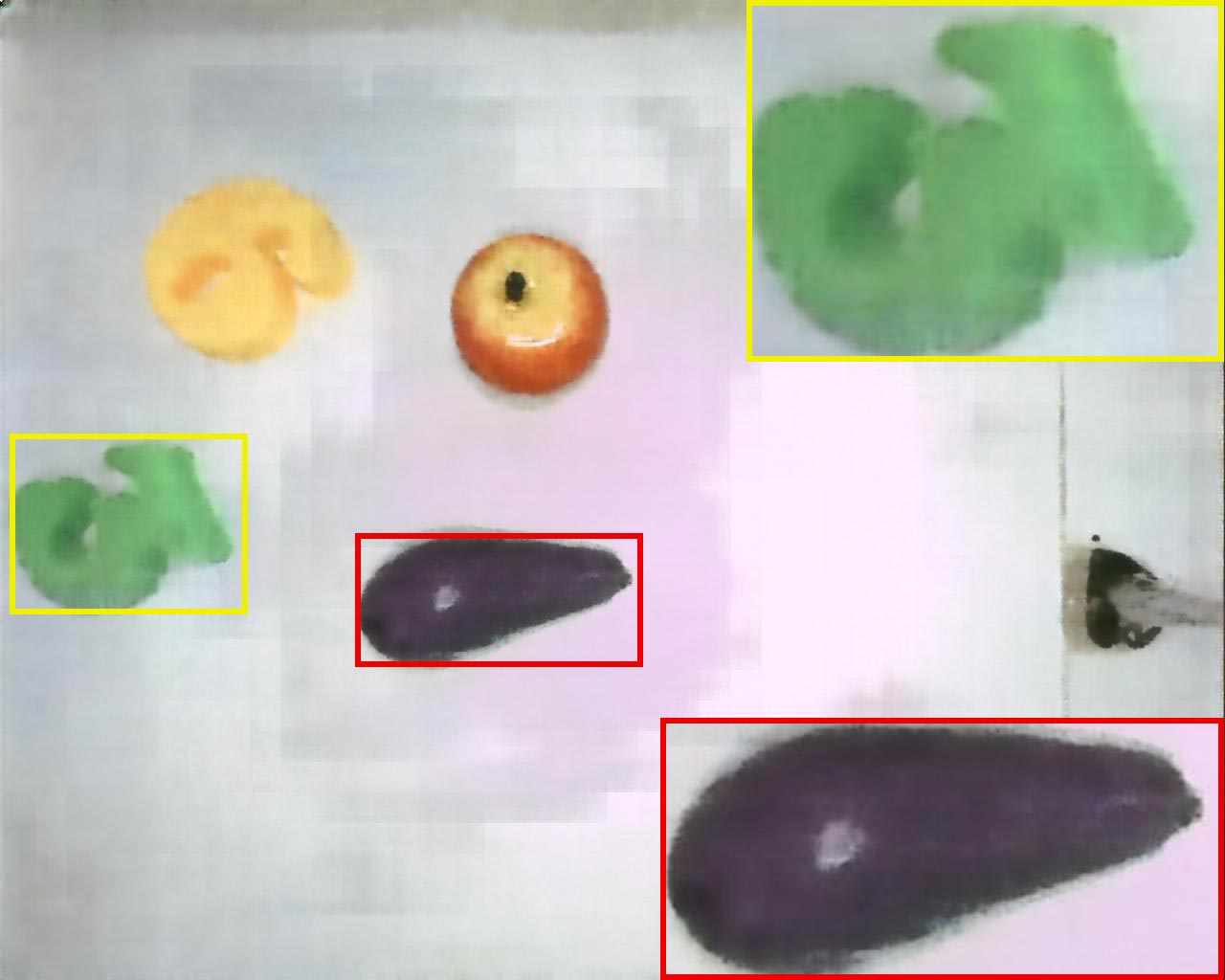}{(g) SID\cite{chen2018learning} }
\imgStubPDFpage{0.19}{width=0.99\linewidth}{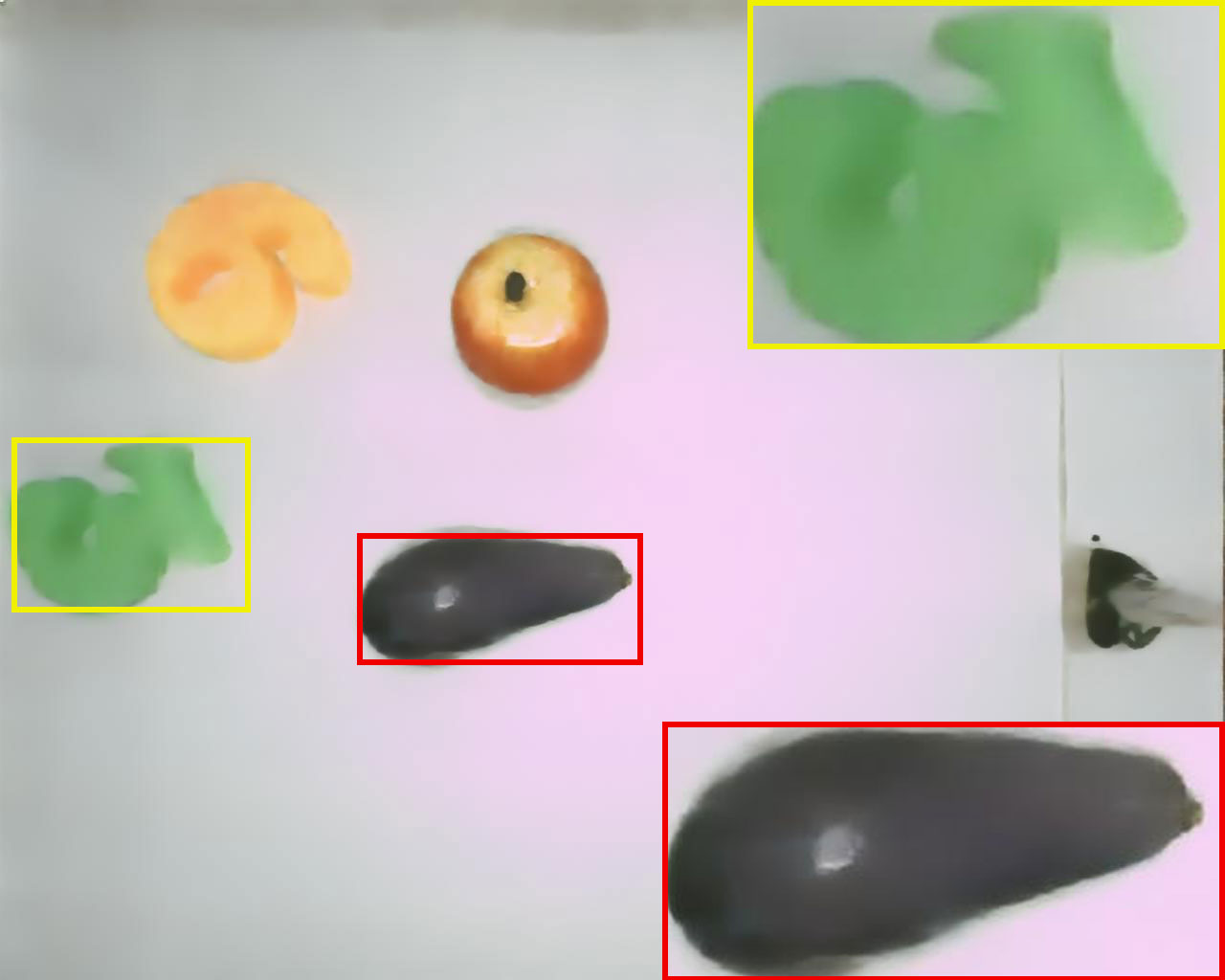}{(h) LDC\cite{xu2020LDC} }
\imgStubPDFpage{0.19}{width=0.99\linewidth}{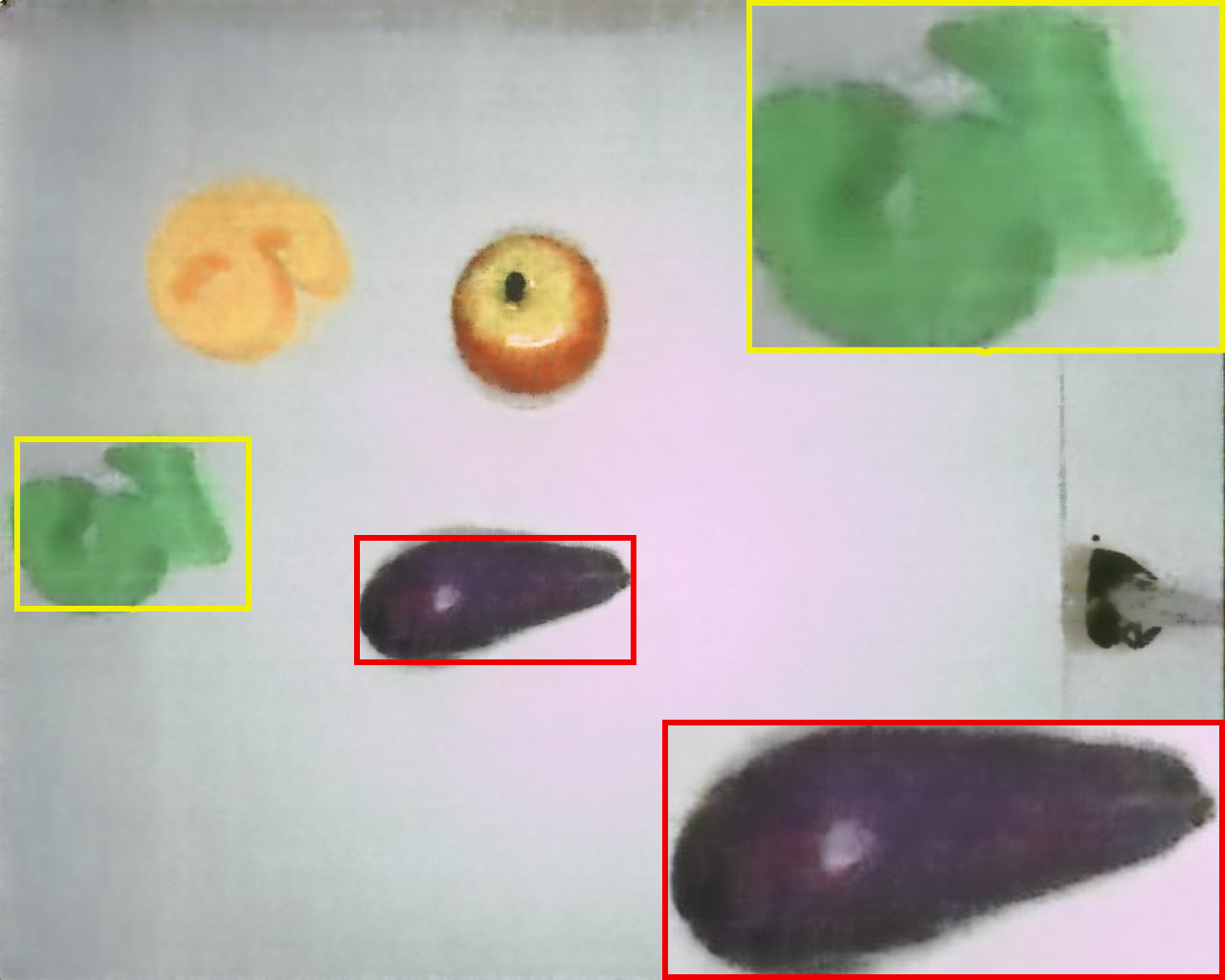}{(i) Ours }
\imgStubPDFpage{0.19}{width=0.99\linewidth}{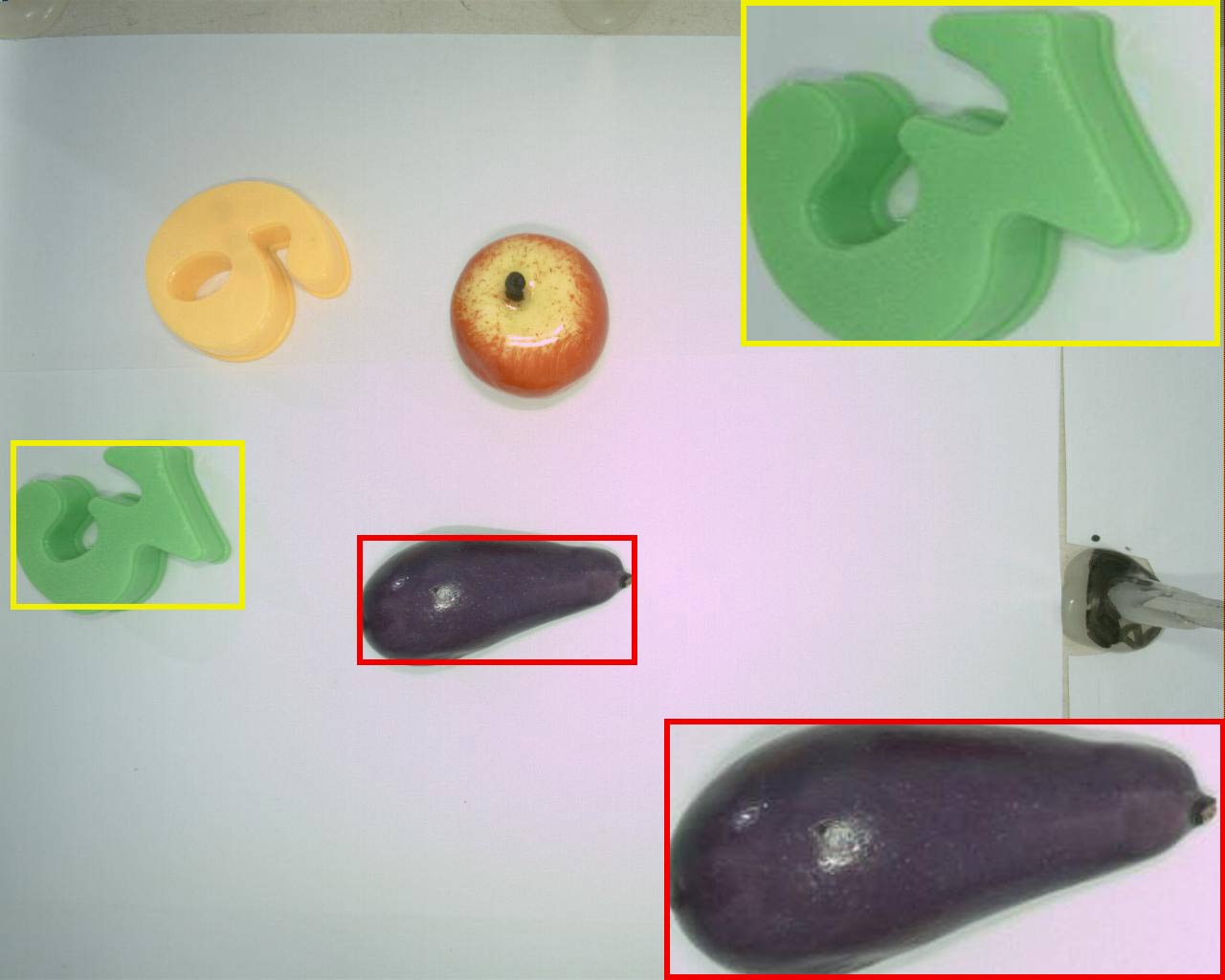}{(j) GT }
\imgStubPDFpage{0.19}{width=0.99\linewidth}{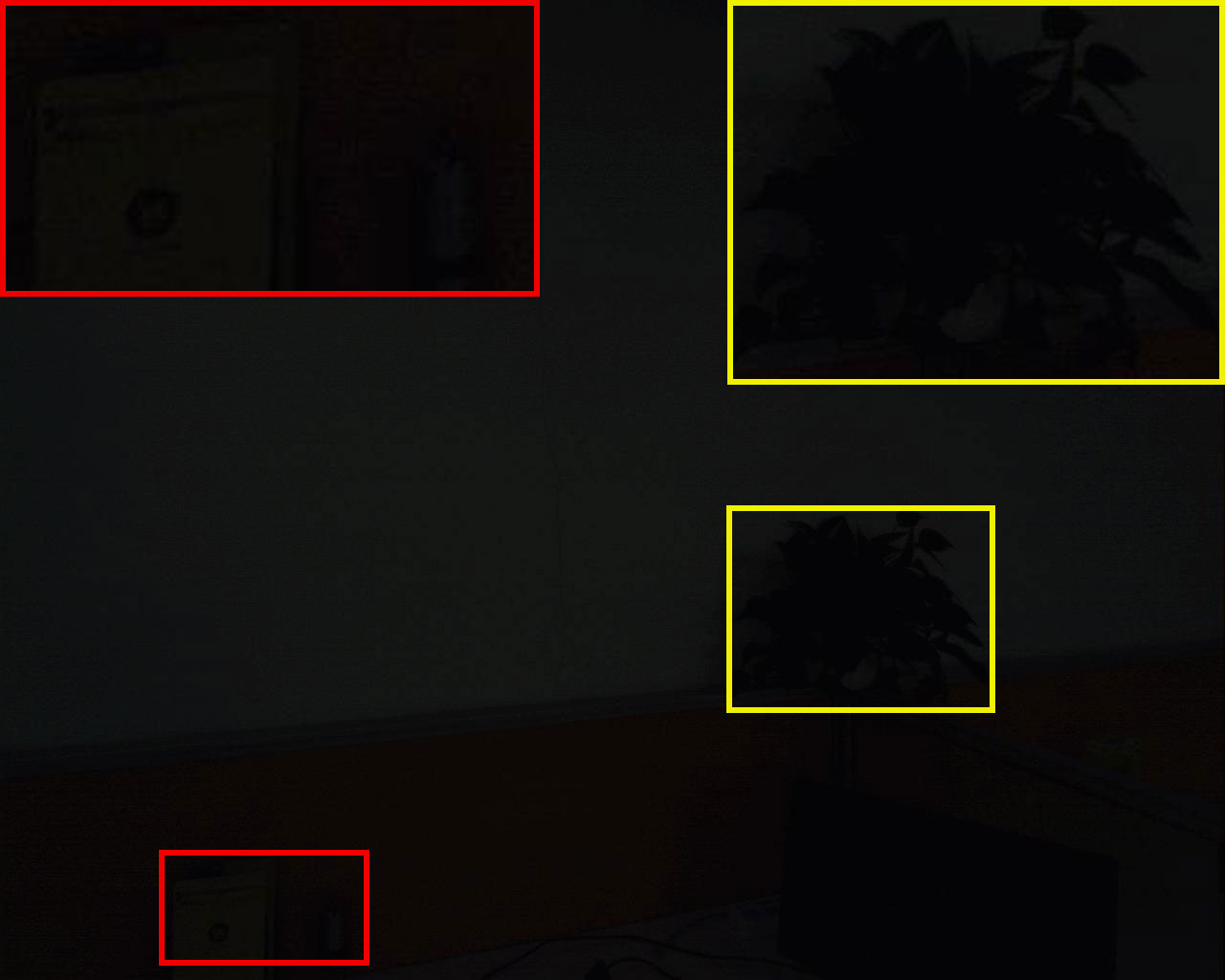}{(A) Input raw }
\imgStubPDFpage{0.19}{width=0.99\linewidth}{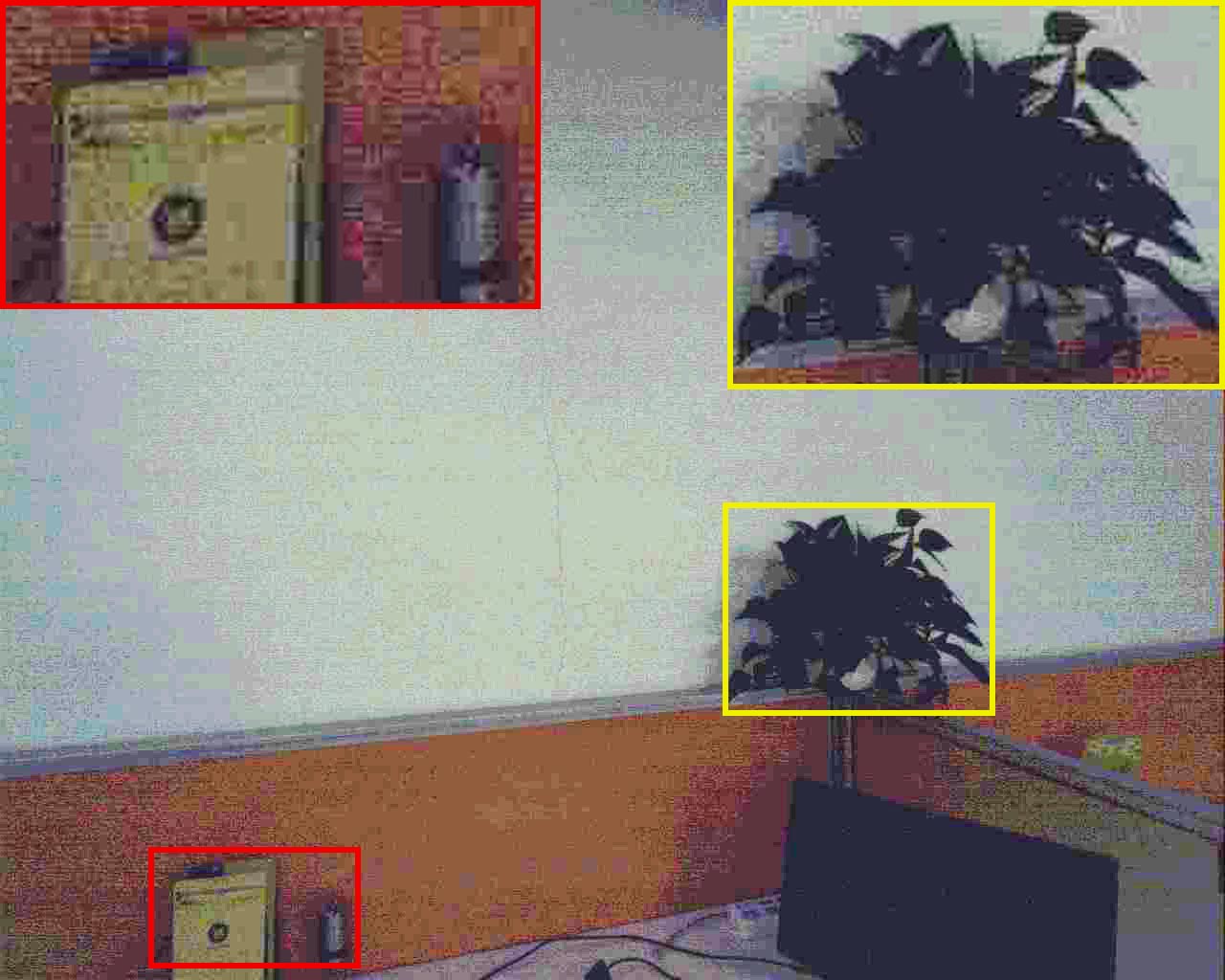}{(B) CSAIE}
\imgStubPDFpage{0.19}{width=0.99\linewidth}{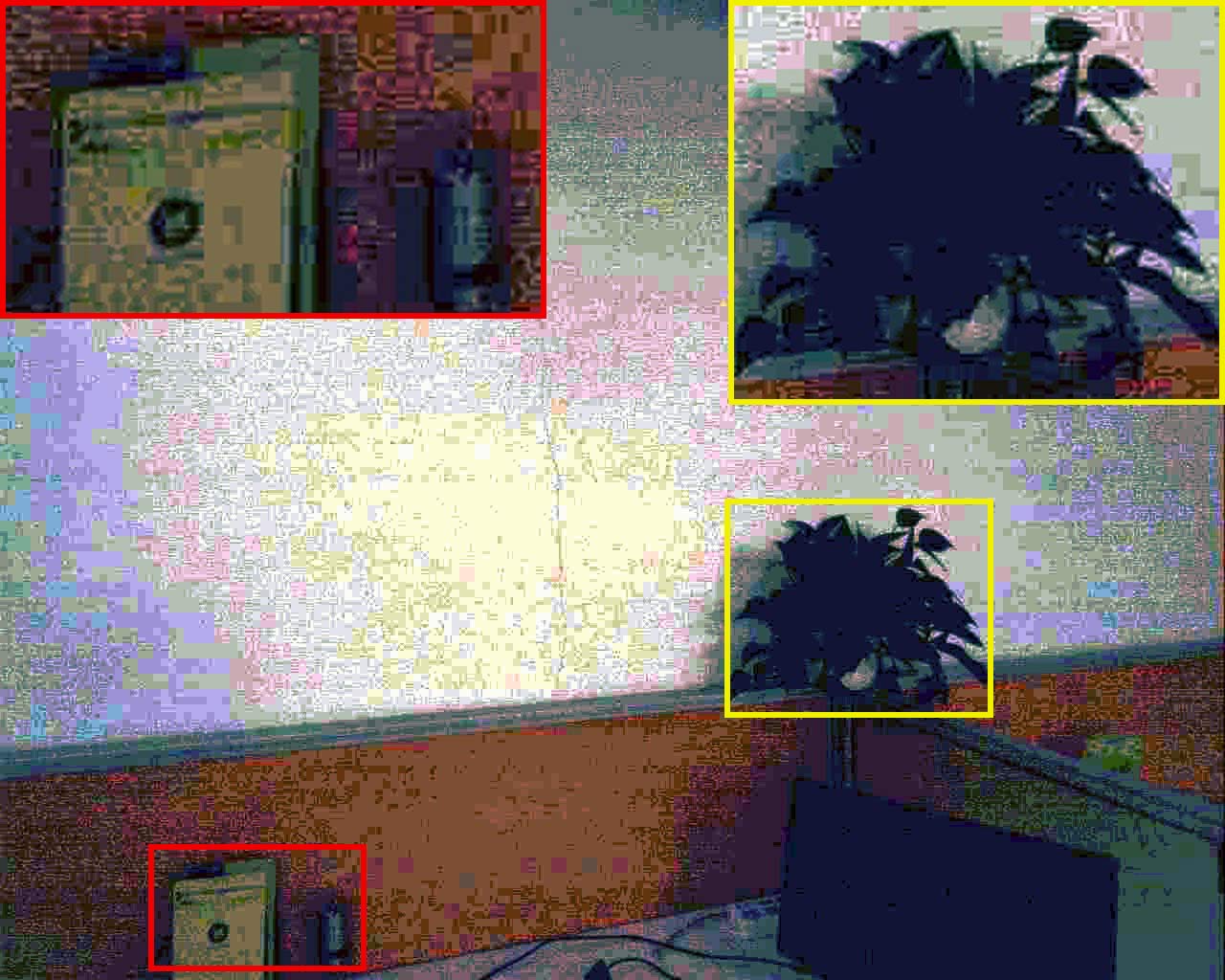}{(C) HE }
\imgStubPDFpage{0.19}{width=0.99\linewidth}{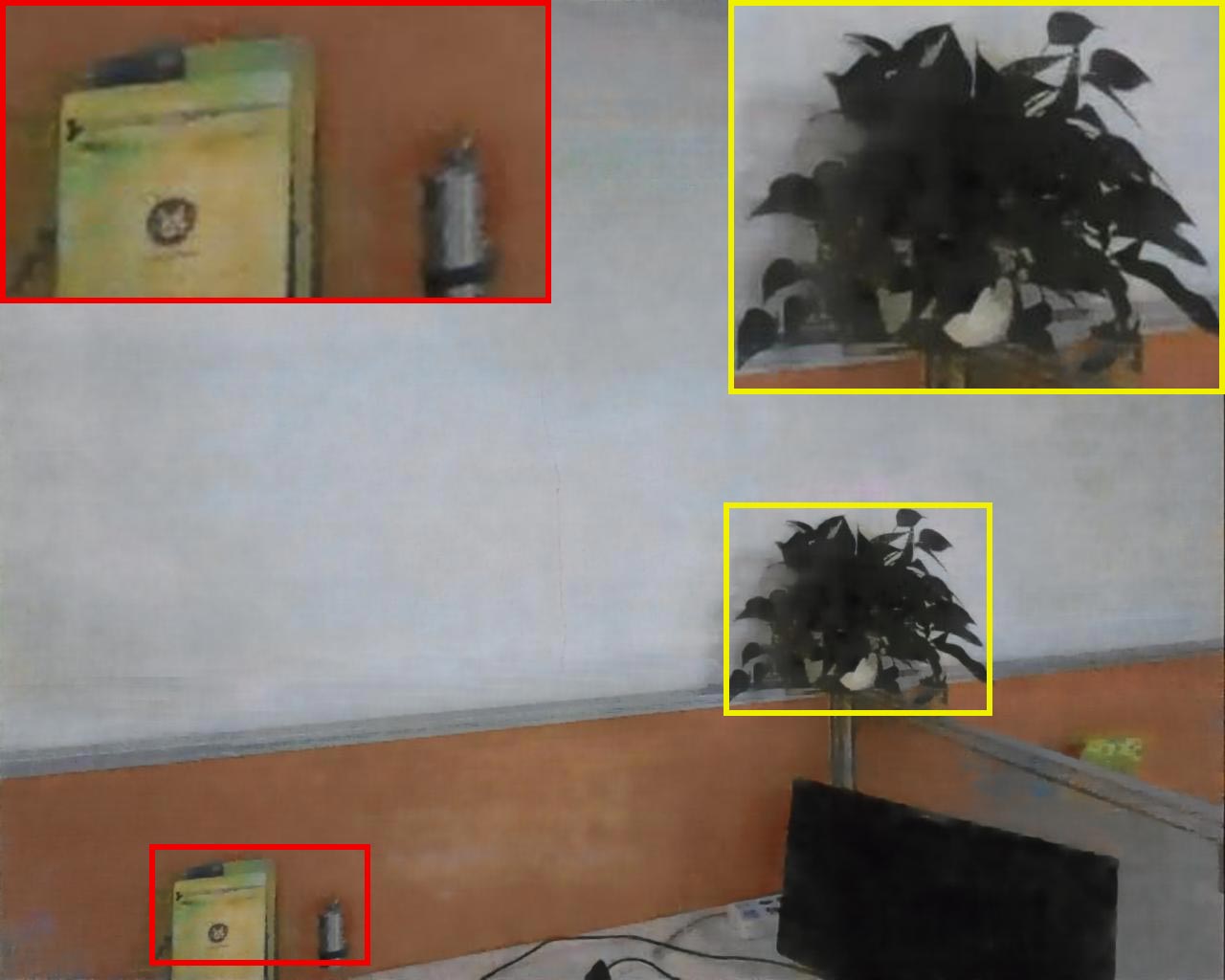}{(D) SGN \cite{gu2019SGN} }
\imgStubPDFpage{0.19}{width=0.99\linewidth}{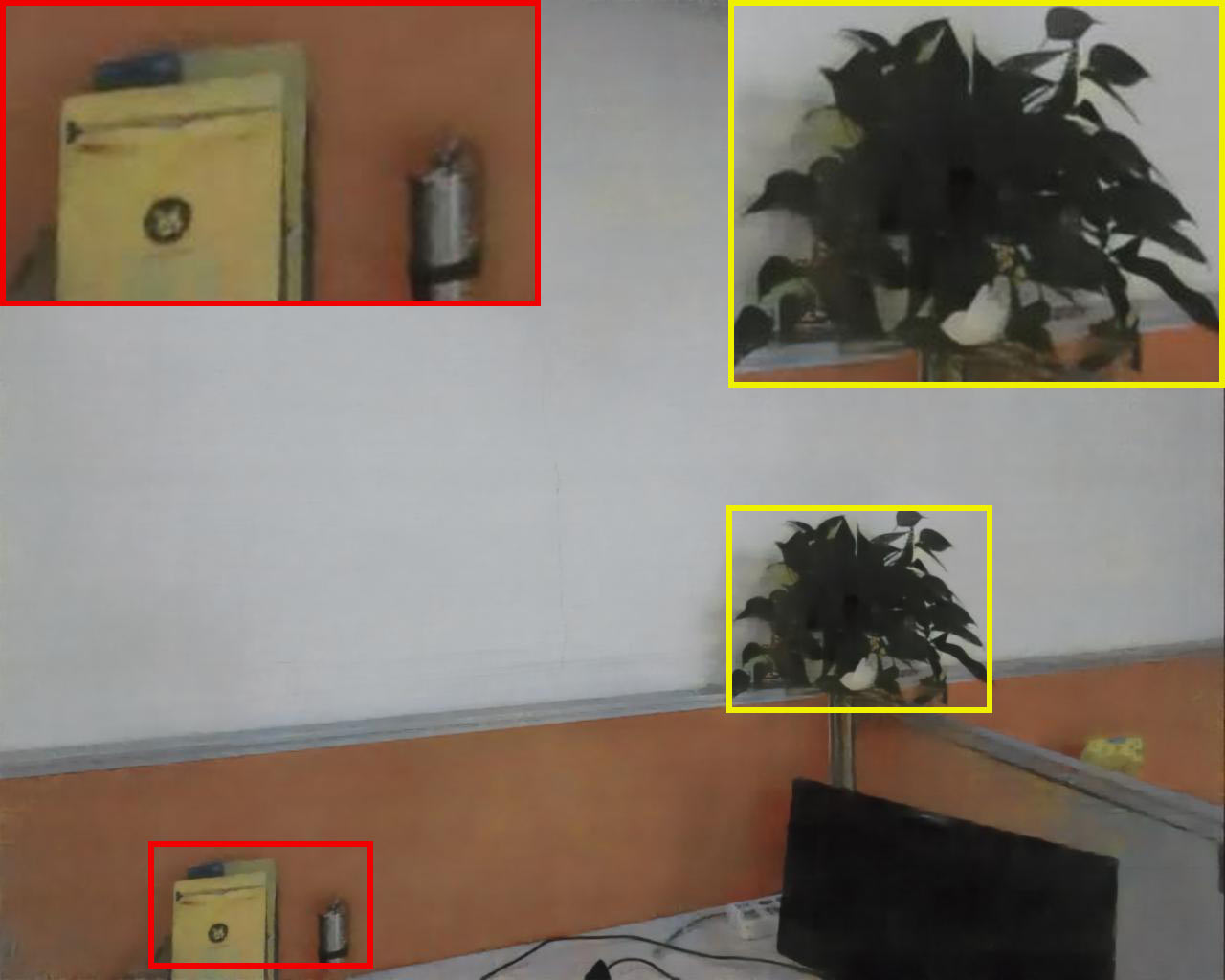}{(E) DID\cite{maharjan2019DID} }
\imgStubPDFpage{0.19}{width=0.99\linewidth}{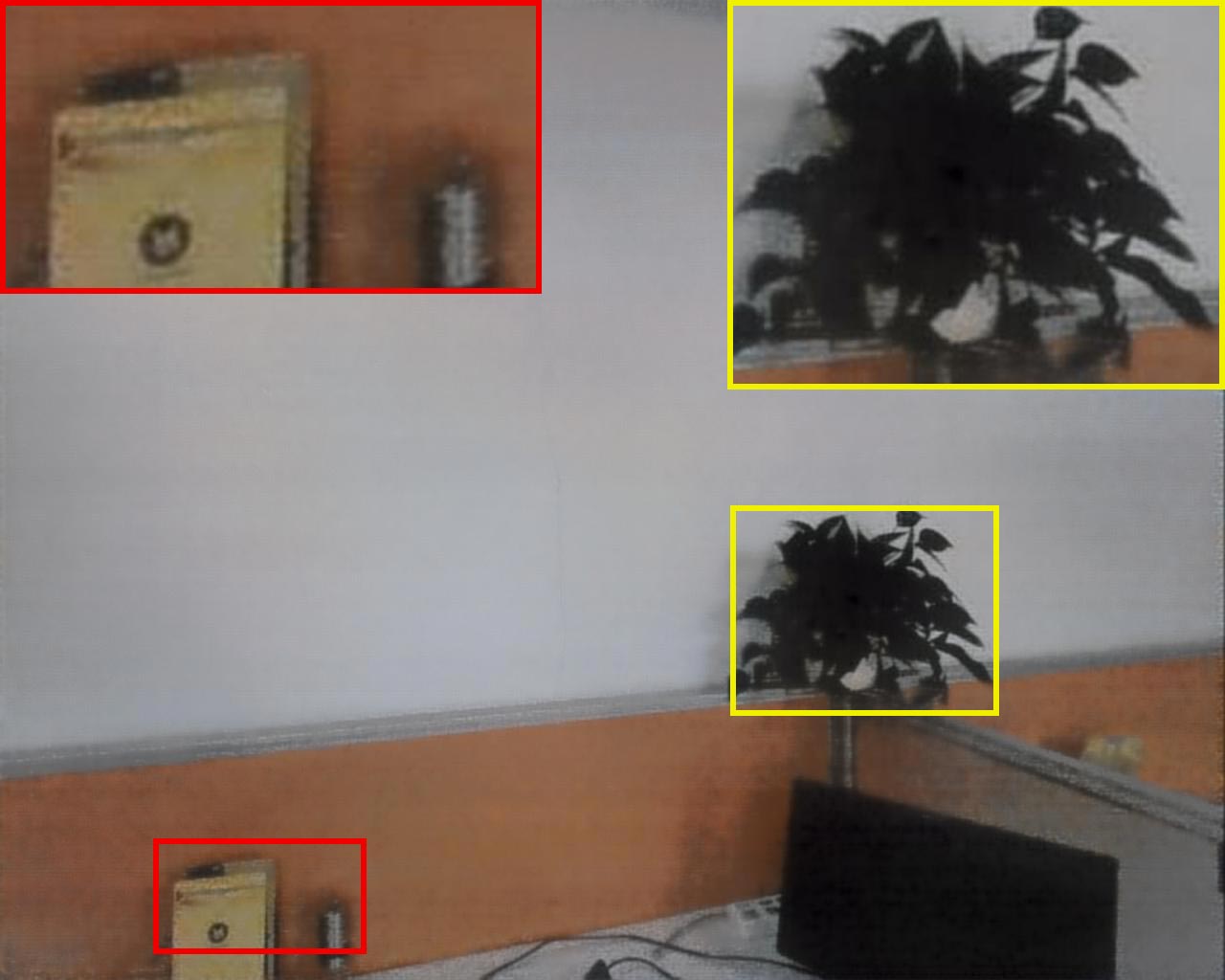}{(F) RED }
\imgStubPDFpage{0.19}{width=0.99\linewidth}{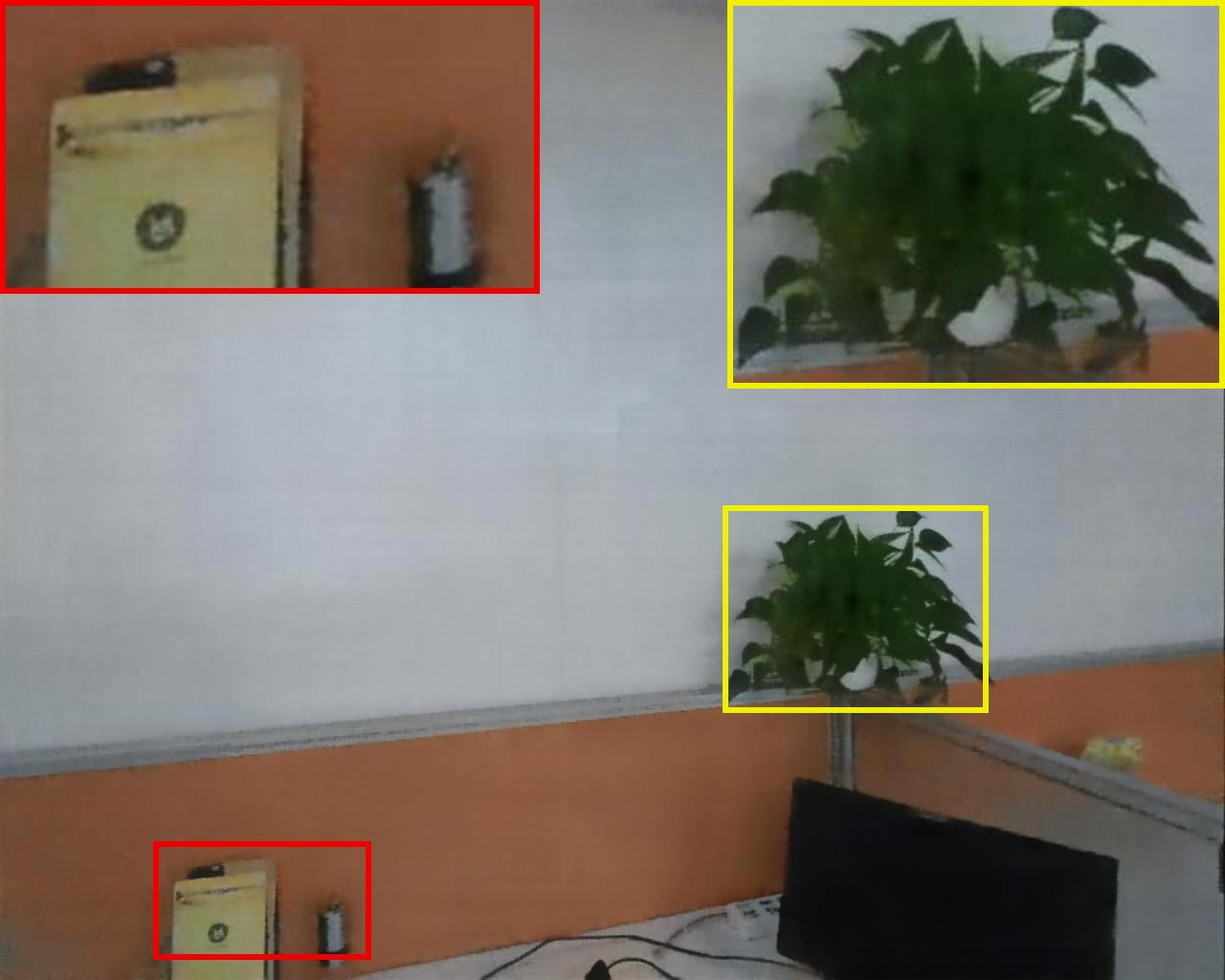}{(G) SID\cite{chen2018learning} }
\imgStubPDFpage{0.19}{width=0.99\linewidth}{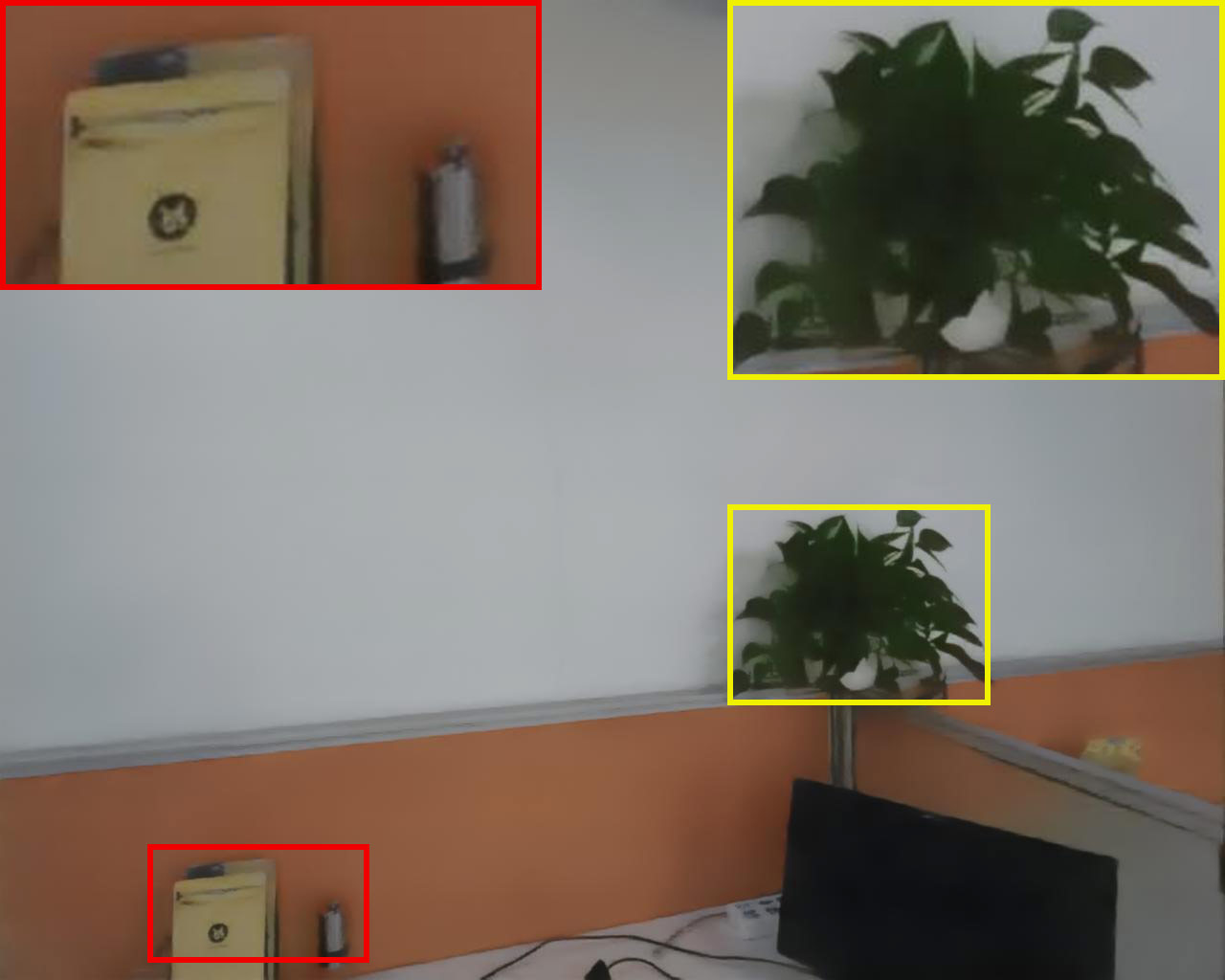}{(H) LDC\cite{xu2020LDC} }
\imgStubPDFpage{0.19}{width=0.99\linewidth}{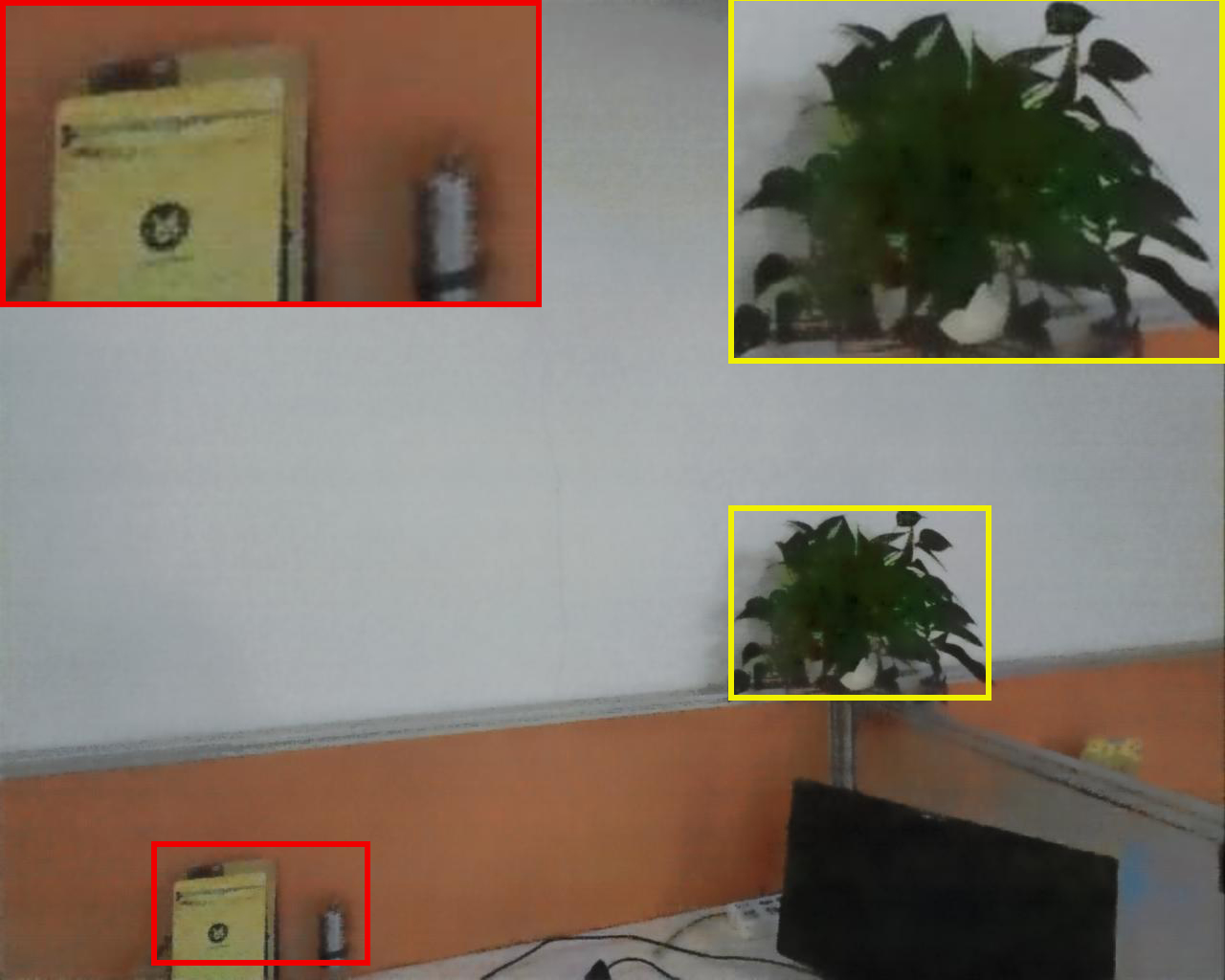}{(I) Ours }
\imgStubPDFpage{0.19}{width=0.99\linewidth}{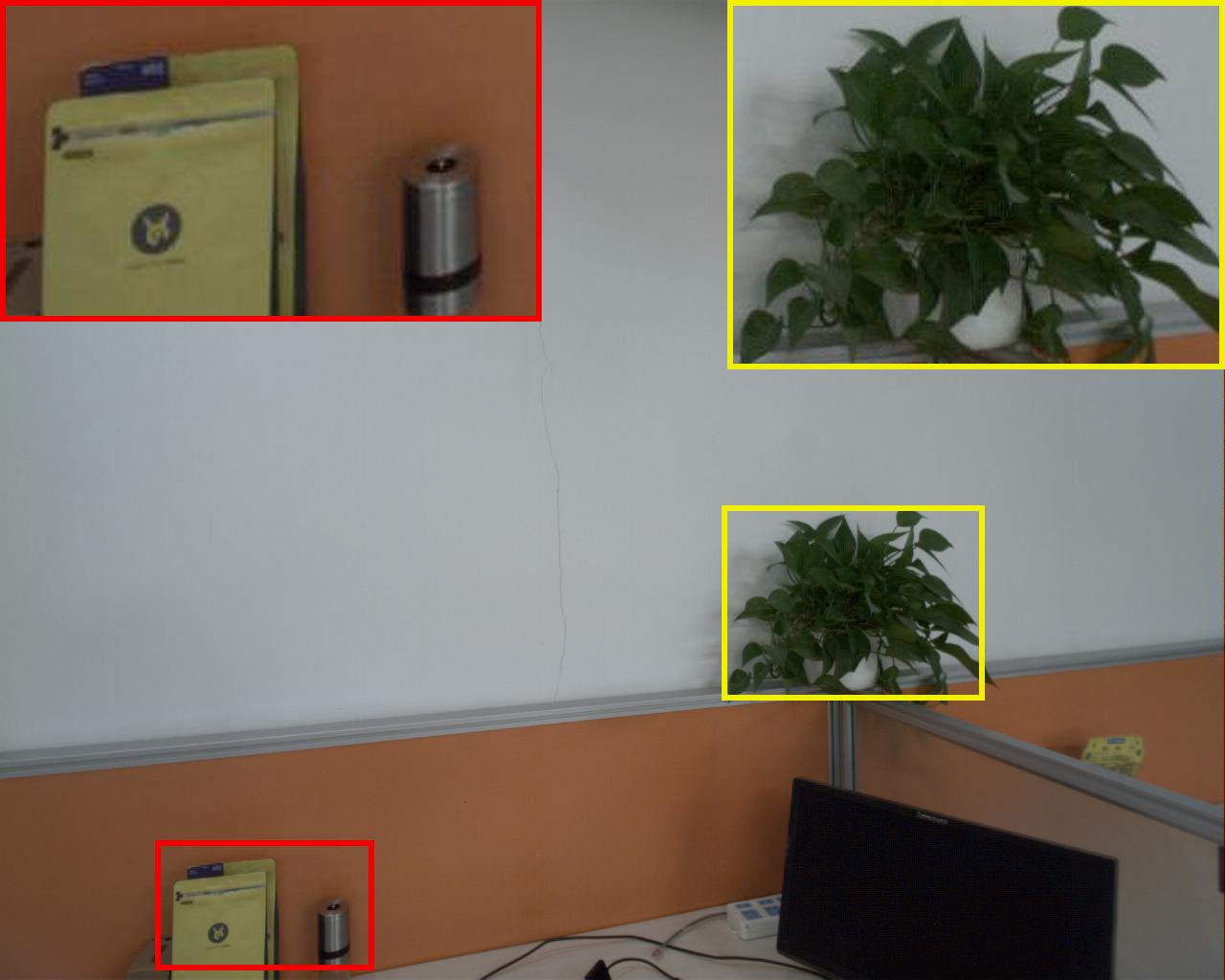}{(J) GT }

\caption{Visual results of state-of-the-art methods and ours on low-light images RAW in our dataset. The larger boxes show the zoom-in version of the regions in the smaller boxes of the same color. \wanyan{The 'CSAIE' means 'Commercial Software Automatic Image Enhancement'}.
 \label{Figure:visuallyComparisonwithsotaraw1}}
\end{figure*}
\subsection{Training}
By default, we pre-process the input images similarly to \cite{chen2018learning} where images' pixel values are amplified with predefined ratios followed by a pack raw operation. We incorporate the CA layer\cite{hu2018squeeze} to bridge the domain gap between features from monochrome and colored raw images. The whole system is trained jointly with L1 loss to directly output the corresponding long-exposure monochrome and sRGB images.

The dataset is split into train and test sets without overlapping by the ratio of 9:1. The input patches are randomly cropped from the original images with 512 × 512. In the case of raw image input, the RGGB pixel position is carefully preserved in the cropping process. \xingbo{We implement our model with Pytorch 1.7 on the RTX 3090 GPU platform, and we train the networks from scratch using the Adam\cite{kingma2014adam} optimizer. The learning rate was set to $10^{-4}$ and $10^{-5}$ after converging, and the weight-decay was set to 0.}

\section{Experiments and results}
In this section, we present a comprehensive performance evaluation of the proposed low-light image enhancement system. To measure the performance, we evaluate the system performance in terms of peak signal-to-noise ratio (PSNR) and structural similarity (SSIM). For PSNR and SSIM, the higher value means the better similarity between output image and ground truth. 
\subsection{Comparison with State-of-the-Arts}

\Stress{Qualitative comparison.} We first visually compare the results of the proposed method with other state-of-the-art deep learning-based image enhancement methods, including SID\cite{chen2018learning}, DID\cite{maharjan2019DID}, SGN\cite{gu2019SGN}, LDC\cite{xu2020LDC}, and RED\cite{RealTimeDarkImageRestorationCvpr2021}. In addition, the traditional histogram equalization (HE) approach and a \wanyan{Commercial Software Automatic Image Enhancement (CSAIE)} method are also included in the comparison. Figure \ref{Figure:visuallyComparisonwithsotaraw1} shows the results of different methods on two low-light images  (see more results in supplementary). 

As indicated by Figure \ref{Figure:visuallyComparisonwithsotaraw1}, our method can achieve better enhancement and denoising visual performance. Specifically, checkerboard artifacts are usually found on SID for images with white background. This is because of the usage of up-sampling layers in the model. Foggy artifacts are usually observed on SGN; color distortions also are found on SGN, DID, and RED, as are shown in Figure \ref{Figure:visuallyComparisonwithsotaraw1} (A-J), where the green plant enclosed by the yellow box becomes black after restoring by SGN, DID, and RED. Compared to LDC, our methods can preserve more details as \textit{over-smoothing} is usually found on LDC. Note that \textit{over-smoothing may be more visual appealing, but details will be lost, for example, the wall crack becomes invisible on LDC as shown in Figure \ref{Figure:visuallyComparisonwithsotaraw1} (H-I)}. In a nutshell, Figure \ref{Figure:visuallyComparisonwithsotaraw1} demonstrates the satisfying visual performance achieved by our method, with fewer artifacts but more convincing restoration.

\begin{table}
\centering
\caption{Comparison with SOTA.\label{table.SOTA}}
\resizebox{\linewidth}{!}{%
\begin{tabular}{c|c|c|c|c} 
\hline
 & \multicolumn{2}{c|}{MCR Dataset} & \multicolumn{2}{c}{SID Dataset} \\ 
\hline
 & PSNR (dB) & SSIM & PSNR (dB) & SSIM\\ 
\hline
RED\cite{RealTimeDarkImageRestorationCvpr2021} \emph{(21,CVPR)} & 25.74 & 0.851 & 28.66 & 0.790 \\ 
\hline
SGN\cite{gu2019SGN} \emph{(19,ICCV)} & 26.29 & 0.882 & 28.91 & 0.789 \\ 
\hline
DID\cite{maharjan2019DID} \emph{(19,ICME)} & 26.16 & 0.888 & 28.41 & 0.780 \\ 
\hline
SID\cite{chen2018learning} \emph{(18,CVPR)} & 29.00 & 0.906 & 28.88 & 0.787 \\ 
\hline
LDC\cite{xu2020LDC} \emph{(20,CVPR)} & 29.36 & 0.904 & 29.56 & \textbf{0.799} \\ 
\hline
Ours & \textbf{31.69} & \textbf{0.908} & \textbf{29.65} & \underline{0.797} \\
\hline
\end{tabular}
}
\end{table}
\Stress{Quantitative comparison.} A quantitative comparison against the state-of-the-art enhancement methods has also been performed. For a fair comparison, SID\cite{chen2018learning}, DID\cite{maharjan2019DID}, SGN\cite{gu2019SGN}, LDC\cite{xu2020LDC}, and RED\cite{RealTimeDarkImageRestorationCvpr2021} were trained on the MCR dataset. 

As Table \ref{table.SOTA} shows, our proposed method outperforms its counterparts by a large margin. Specifically, our method can achieve a PSNR of 31.69dB on MCR dataset, which is 7.9\% higher than the second-best method, i.e., the LDC\cite{xu2020LDC}. Our method can also achieve an SSIM of 0.908, which is the highest among all compared methods. 

Compared to other methods, we incorporate the extra monochrome information into the processing pipeline, hence state-of-the-art performance can be achieved. As shown from the first two data rows in Table \ref{table.SOTA}, both RED\cite{RealTimeDarkImageRestorationCvpr2021} and SGN\cite{gu2019SGN} can only achieve a PSNR of around 26dB. Both RED and SGN aim at reducing the computational cost and improving efficiency. Hence it is reasonable to observe the performance degradation. The result on DID\cite{maharjan2019DID} from Table \ref{table.SOTA} suggests that replacing U-net with residual learning can not achieve superior performance on our dataset. 

On the MCR dataset, SID\cite{chen2018learning} achieves only a PSNR of 29.00dB. The checkerboard artifact may be the reason. From Table \ref{table.SOTA}, we observe that LDC\cite{xu2020LDC} achieves the second-best performance. This is because they are based on a frequency-based decomposition and enhancement model, which can better restore the noisy image and avoid noise amplification. We also train our model on the modified SID dataset to further validate our method for a fair comparison. The performance results are shown in the SID column in Table \ref{table.SOTA}. As the results suggest, our method also outperforms all its counterparts. Specifically, our method can achieve a PSNR of 29.65dB, which is around 0.1dB higher than LDC, while the SSIM can achieve similar performance. 

Other methods including SID, DID, SGN, and RED, can only achieve a PSNR around 28dB.
In summary, the result shows that our model is more effective in enhancing low-light images with noise. The performance of most existing methods is upper bounded by the information contained in the raw data. In our proposed pipeline, we further extend the upper bound by considering the monochrome domain. Hence better performance can be achieved.

\subsection{Ablation study}
In this subsection, we provide several ablation studies for the proposed system to better demonstrate the effectiveness of each module of our system. 

\begin{figure*}
\imgStubPDFpage{0.24}{width=0.99\linewidth}{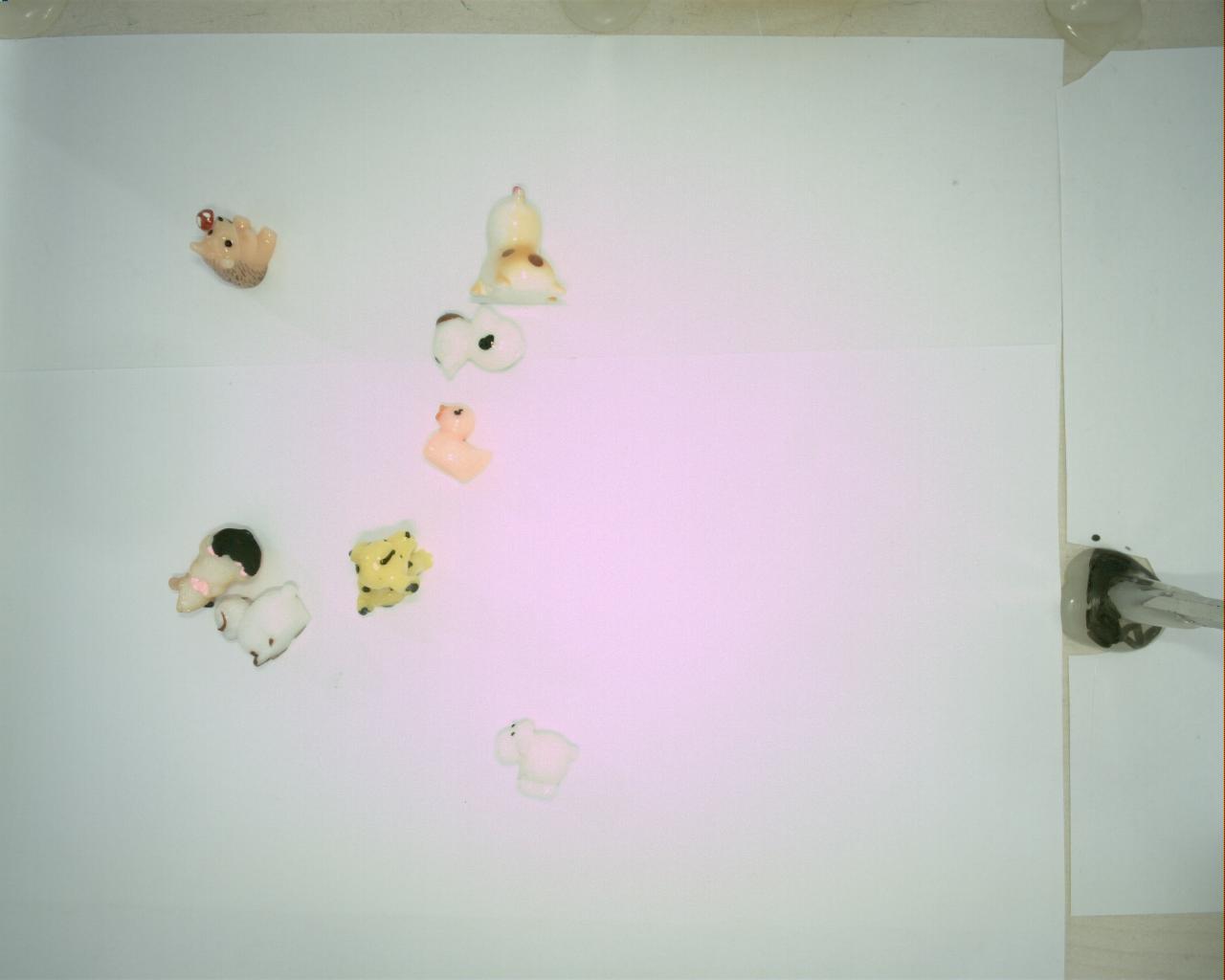}{(a) GT}
\imgStubPDFpage{0.24}{width=0.99\linewidth}{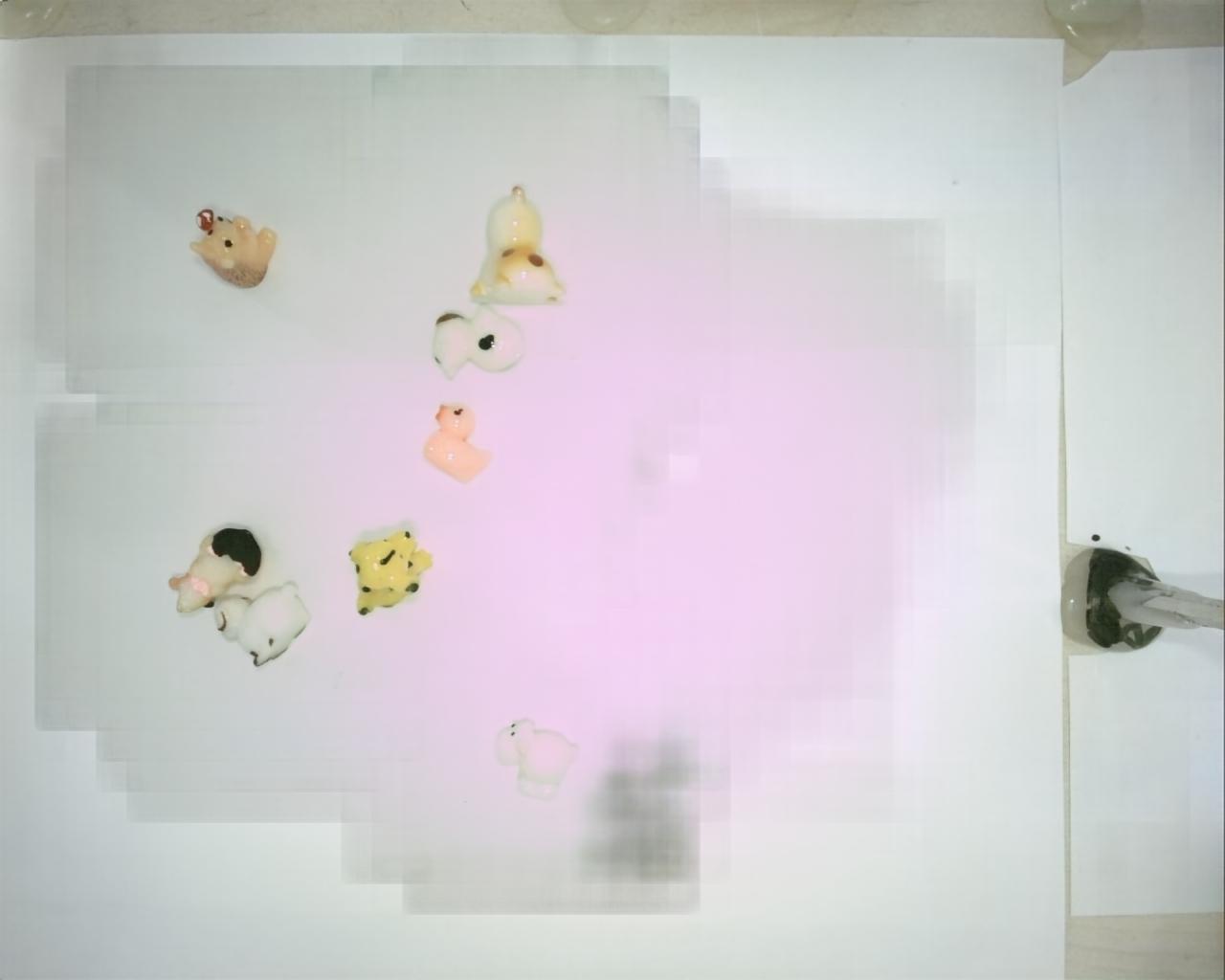}{(b) SID\cite{chen2018learning}}
\imgStubPDFpage{0.24}{width=0.99\linewidth}{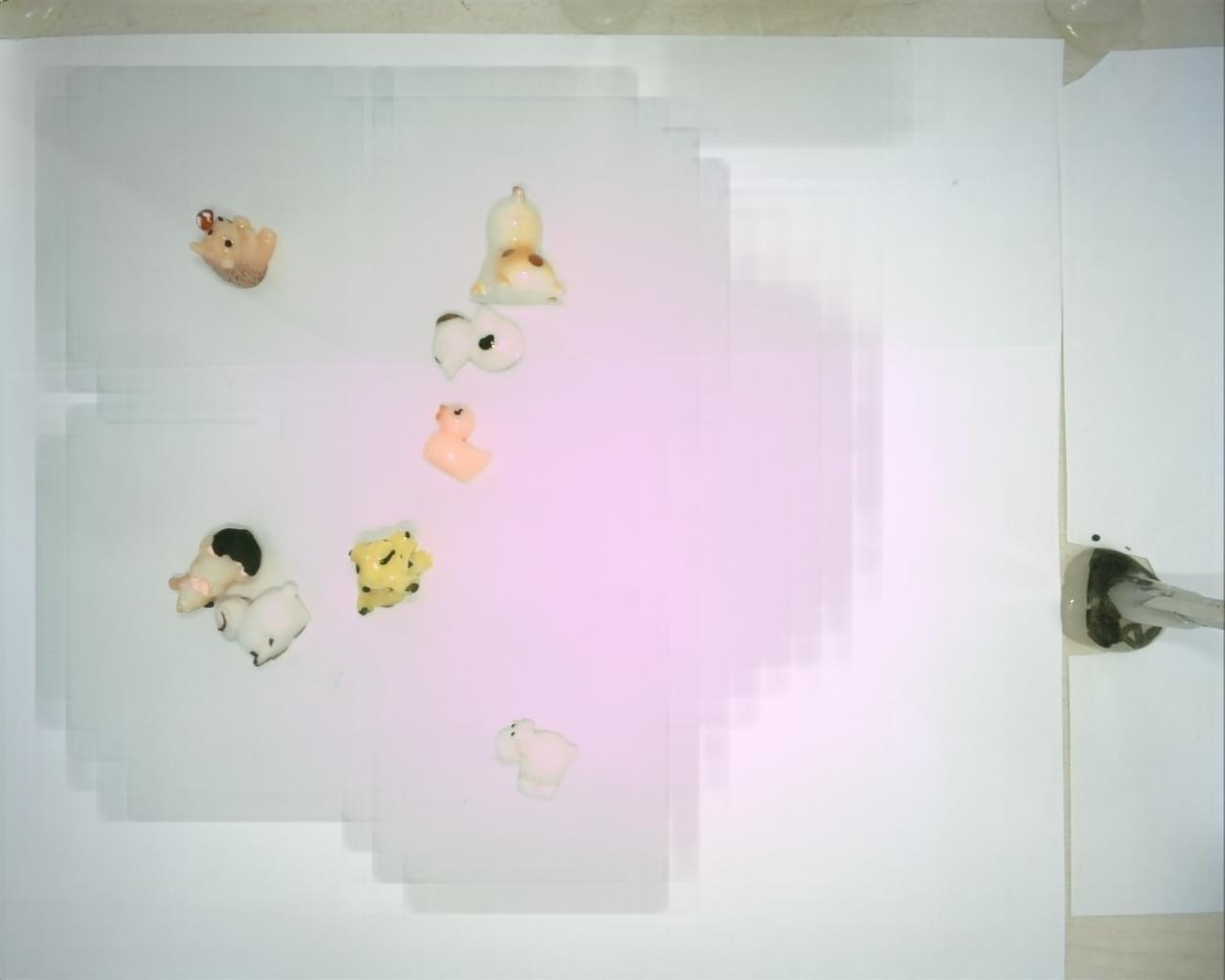}{(c) Ours w/o CA\cite{hu2018squeeze}}
\imgStubPDFpage{0.24}{width=0.99\linewidth}{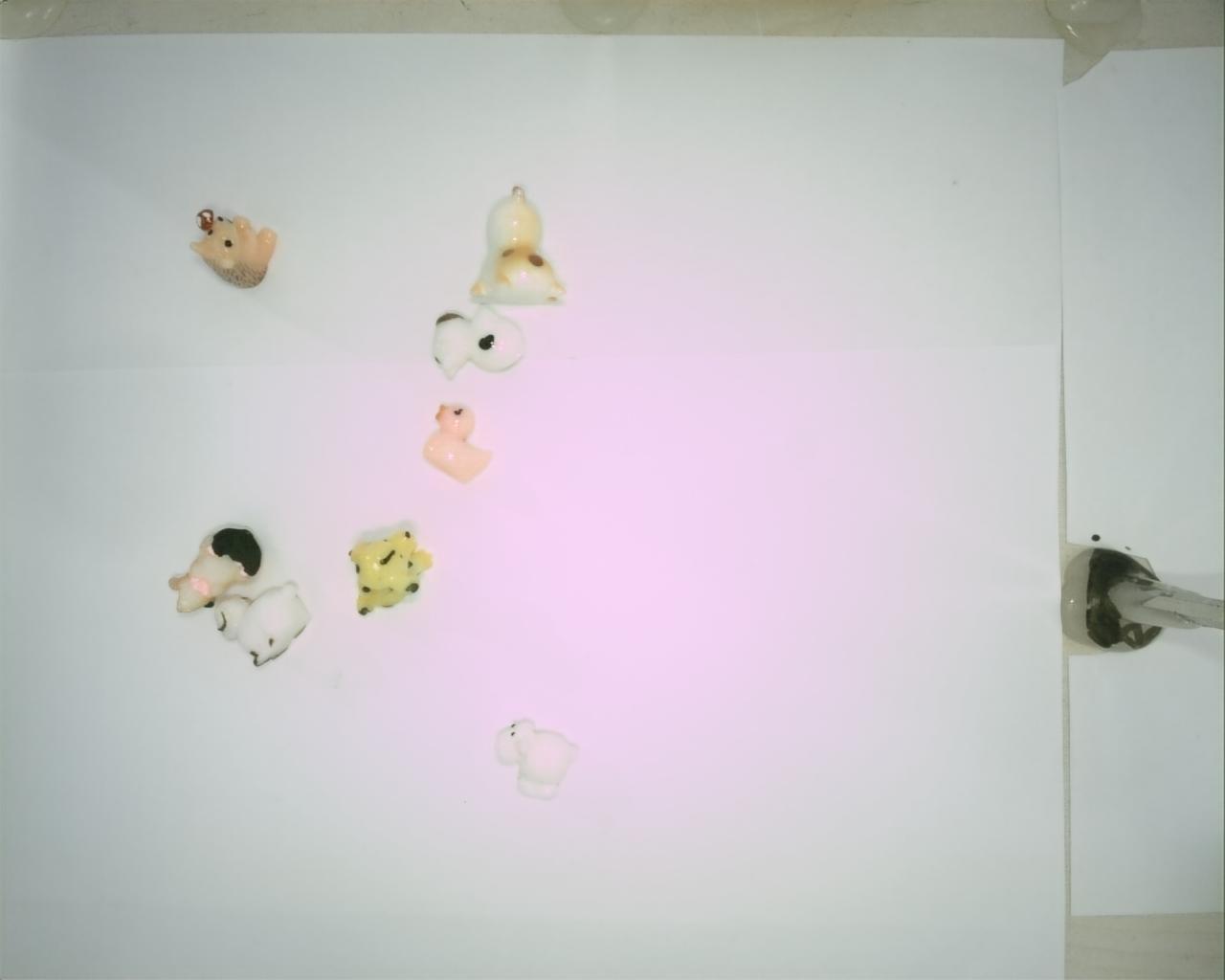}{(d) Ours with CA\cite{hu2018squeeze}}
\caption{Visual demonstration of checkerboard artifacts under different settings.\label{Figure:checkerboard}}
\end{figure*}

Checkerboard artifacts are found in our preliminary exploration stage, especially for images with white backgrounds. To eliminate checkerboard artifacts, we incorporate the CA layer\cite{hu2018squeeze} in the DBLE module. In this ablation study, we first remove the CA layer in the DBLE module to demonstrate the checkerboard artifacts elimination and performance upgrading. Besides, we also train an original SID\cite{chen2018learning} network on our dataset to show the visual effect of the checkerboard artifacts of U-net. The restored images from SID, DBLE w/o CA layer, and DBLE with CA layer are shown in Figure \ref{Figure:checkerboard}. It is observed that checkerboard artifacts can be perfectly avoided by introducing the CA layer. Besides, as the quantitative results shown in Table \ref{table.ablationstudy}, CA layer can boost the image enhancement performance as the PSNR increases to 31.69dB compared with its counterpart 29.23dB. 

We also train the model to learn the ratio directly instead of amplifying image pixel values with predefined ratios. Hence we train a model without amplifying the input raw images with the predefined ratio. As a result, as shown in Table \ref{table.ablationstudy}, such a model can still achieve comparable performance, with only a slight decrease in PSNR and SSIM.

As suggested by \cite{chen2018learning}, we change the packraw-based input into original one channel raw images. As shown in the row of baseline without packraw in Table \ref{table.ablationstudy}, PSNR and SSIM degradation is observed. We argue that the packing of raw can assist the model to better process the color information. 

The change of loss function from L1 to L2 can not achieve better performance, as shown in Table \ref{table.ablationstudy}. We also try to change the input raw into sRGB format. The result in sRGB row from Table \ref{table.ablationstudy} shows a significant performance drop, which is consistent with other works \cite{chen2018learning,xu2020LDC}. 

The DBF module plays a key role in our system in generating the monochrome images, which assist the DBLE module in restoring the low-light images into monitor-ready sRGB images. 
We also explore the performance of a model without DBF module and the monochrome branch. As the result in Table \ref{table.ablationstudy} shows, the performance drops to 29.99dB/0.883 in terms of PSNR/SSIM when the DBF module is removed, hence providing a solid validation of the DBF's effectiveness. 

\begin{table}
\centering
\caption{Ablation study on the MCR dataset.\label{table.ablationstudy}}
\resizebox{\linewidth}{!}{%
\begin{tabular}{l|c|c|c|c} 
\hline
 & \multicolumn{2}{c|}{DBF} & \multicolumn{2}{c}{DBLE} \\ 
\hline
 & PSNR (dB) & SSIM & PSNR (dB) & SSIM \\ 
\hline
Baseline & \textbf{21.0607} & \textbf{0.8254} & \textbf{31.6905} & \textbf{0.9083} \\ 
\hline
Baseline wo CA\cite{hu2018squeeze} & 20.2673 & 0.7948 & 29.2350 & 0.8732 \\ 
\hline
Baseline wo ratio & 19.8978 & 0.7868 & 29.3528 & 0.8878 \\ 
\hline
Baseline wo packraw & 20.7846 & 0.8034 & 28.8728 & 0.8657 \\ 
\hline
Baseline l1$\rightarrow$l2 & 20.4587 & 0.8016 & 30.2359 & 0.8974 \\ 
\hline
Baseline w/o DBF & - & - & 29.9946 & 0.8839 \\ 
\hline
Baseline raw$\rightarrow$ sRGB & 18.2369 & 0.7625 & 27.3521 & 0.8295 \\ 
\hline

\end{tabular}
}
\end{table}

\wanyan{
\section{Limitations and future work}
There are various aspects to improve in the future. 
The cameras we adopted in this work can only output 8 bits raw images, the 16 bits cameras will be used to collect data in the future to cover more diverse scenes and objects. Besides, the network complexity needs to be more light-weighted to deploy the proposed system in the real world. Additionally, extending the proposed work to videos will also be one future direction. We hope the work presented in this paper can provide preliminary explorations for the low-light image enhancement research in community and industry. 
When it comes to some extremely dark images on our MCR Dataset, the existing low-light image enhancement algorithms (SID\cite{chen2018learning}, LDC\cite{xu2020LDC}, and ours) show unsatisfying results sometimes. The restored images usually lost the high-frequency edge information compared to the ground truth image and became blurred (see in supplementary). Extremely dark settings sometimes yield quite weak signals in each color channel, leading to those color artifacts that commonly exist in both SoTA and ours methods and require further study.
}

\section{Conclusion}
Removing the Bayer filter allows more photos to be captured by the sensor. Motivated by this fact, this work proposes an end-to-end fully convolutional network consisting of a DBF module and a dual branch low-light enhancement module to achieve low-light image enhancement on a single colored camera system. The DBF module is devised to predict the corresponding monochrome raw image from the color camera raw data input. The DBLE is designed to restore the low-light raw images based on the raw input and the DBF predicted monochrome raw images. DBLE treats the colored raw and monochrome raw separately by using a dual branch network architecture. In the DBLE up-sampling stream, features from both monochrome raw and colored raw are fused together and a channel-wise attention is applied on the fused features.

We also propose a Mono-Colored Raw paired dataset (MCR) which includes color and monochrome raw image pairs collected by a color camera with Bayer-Filter and a monochrome camera without Bayer-Filter. The dataset is collected in various scenes, and each colored raw image has a corresponding monochrome raw image captured with the same exposure settings. To better show our superiority, the SID dataset is also adopted in the evaluation. Gray image is generated from the corresponding ground truth color image in the SID dataset to serve as the monochrome image. Subsequently, a model is trained on the modified dataset to verify the performance. 

Our experiments prove that significant performance can be achieved by leveraging raw sensor data and data-driven learning. Our method can overcome the checkerboard artifact which is found on U-net, while preserving the visual quality. Our quantitative experiments indicate that our methods can achieve the state-of-the-art performance: a PSNR of 31.69dB on our own dataset, and 29.65dB on the SID dataset.

{\small
\bibliographystyle{ieee_fullname}
\bibliography{egbib}
}

\end{document}